\documentclass[floatfix,prc,twocolumn,amsmath,amssymb]{revtex4}

\usepackage{graphicx}

\newcommand{\be}{\begin{eqnarray}}
\newcommand{\ee}{\end{eqnarray}}

\def\npa#1#2#3{    { Nucl. Phys. }{\bf A#1},  #3 (#2)}
\def\npb#1#2#3{    { Nucl. Phys. }{\bf B#1},  #3 (#2)}
\def\plb#1#2#3{    { Phys. Lett. }{\bf B#1},  #3 (#2)} 
\def\prd#1#2#3{    { Phys. Rev. }{\bf D#1},  #3 (#2)}

\def\epjc#1#2#3{    { Eur.Phys.J. }{\bf C#1},  #3 (#2)}

\def\epjc#1#2#3{   { Eur. Phys. J. }{\bf C#1},  #3 (#2)}

\begin{document}

\title{An Impact Parameter Dipole Saturation Model}

\author{Henri Kowalski}
\affiliation{%
 Physics Department, Brookhaven National Laboratory, Upton, NY 11973, USA\\ 
Department of Physics, Columbia University, NY 10027, USA\\
Deutsches Elektronen Synchrotron DESY, Hamburg Germany
}
\author{Derek Teaney}
\affiliation{%
 Physics Department, Brookhaven National Laboratory,
                Upton, N.Y. 11973, U.S.A.
}

\date{\today}

\begin{abstract}
We develop a dipole model for HERA DIS data  which incorporates the 
impact parameter distribution of the proton. The
model describes the 
inclusive total $\gamma^*p$ cross-sections as well as the diffractive $J/\psi$
differential cross-sections. We compare the model with  previous
approaches and show that the $t$-distributions are sensitive to
saturation phenomena. We estimate the boundary of the saturation
region and show that it dominates the data in the low-$Q^2$ region
where the total $\gamma^*p$ cross-section exhibits the same universal rise
as hadronic cross-sections.
The model is then
extended to nuclei and shows  good agreement with the
nuclear shadowing data at small-$x$.  Finally, we  estimate 
the saturation scale in nuclei.
\end{abstract}

\maketitle

\section{Introduction}

One of the most important observation of the HERA experiments is the
rapid rise of the  $\gamma^{*} p$ cross-section with the $\gamma^{*}
p$ center of mass energy $W$ in the deep inelastic scattering (DIS) region. 
This observation indicates that over a region of size  $\sim ~1/Q^2$ 
additional gluons are abundantly radiated at small 
$x \approx Q^2/W^2$.
The simultaneous observation of a significant number of  
diffractive events in DIS lead to the dipole saturation model of
Golec-Biernat and W\"usthoff (GBW)~\cite{ref.GBW}. The model
successfully described the data in the transition region
between  low and intermediate  $Q^{2}$ 
and indicated the emergence of saturation phenomena in DIS.
In spite of its compelling success the model has
some obvious shortcomings. 
First, the treatment of QCD evolution 
is only rudimentary.  Second, the dipole  cross-section 
is integrated over the transverse coordinate although
the gluon density is  expected to be a strongly varying
function of the impact parameter.  
The evolution was  improved by incorporating DGLAP
  into the dipole cross-section thereby extending the
model applicability to the  high $Q^2$ region~\cite{ref.BGK}.  

In this paper we  further improve the dipole model by
introducing  the impact parameter of the proton into the
dipole dynamics. 
For the form of the dipole cross-section we adopt the Glauber-Mueller
formula which can be derived under simplifying assumptions ~\cite{ref.GlMu}. 
The transverse profile of the proton can be probed experimentally by
measuring the  momentum transfer $t$ to the  proton in diffractive
events. Thus diffractive and 
inclusive $\gamma^{*} p$ processes form  a tool to determine the gluon
density and saturation dynamics at HERA. 
Whether 
the experimental data are sufficient to reveal the 
rich non-linear dynamics of saturation and the 
Color Glass Condensate~\cite{CGC} 
remains an important phenomenological 
question~\cite{ref.MSM, LevinTotal, LevinJ} which this work addresses.

The Impact Parameter (IP) Saturated dipole model can be extended to
DIS interactions in nuclei in a natural way.  We 
study  nuclear effects and compare them with data on nuclear shadowing. 
Finally we estimate the saturation scale in heavy nuclei.

\section{Dipole description of DIS}

\begin{figure}
\hspace{0.5cm}
\includegraphics[height=6.5cm,width=8.cm]{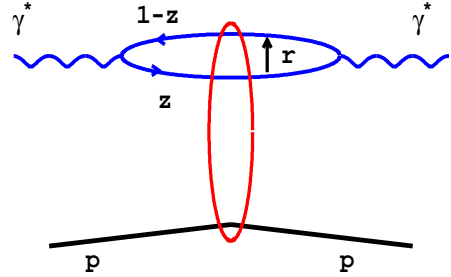}
\caption{\label{fig:fgraqqtot} The interaction of a quark anti-quark dipole with a proton.}
\end{figure}

In 
the dipole picture~\cite{dip1,dip2}, 
the $\gamma^{*}p$ interaction proceeds in
three stages:
First, the 
incoming virtual photon fluctuates into a quark-antiquark pair, 
then the $q\bar{q}$ pair elastically scatters on the proton, and finally
the $q\bar{q}$ pair recombines to form  a virtual photon. This is
shown schematically in Fig.~\ref{fig:fgraqqtot}. 
The amplitude for the complete process is
simply the product of the amplitudes of these three subprocess.
We discuss each in turn. 

The amplitude for the incoming virtual photon (with polarization $\lambda=+,-,0$) to fluctuate into
a quark-antiquark pair (of flavor $f$ with helicities $h,\bar{h}=\pm{1/2}$)   
is given by the 
photon light cone
wave-function
$(\psi_{\gamma,f})_{\lambda}^{h\bar{h}}$ which depends upon 
$(Q^2,z,{\bf r})$.  
Here $z$ denotes the longitudinal momentum fraction of the photon
carried by the quark and  
${\bf r}$ 
the transverse size of the $q\bar{q}$ pair. 
These wave functions are determined from light 
cone perturbation theory to leading order in the fermion 
electromagnetic charge~\cite{ref.BjKS}. 
Similarly, the
amplitude for the $q\bar{q}$ to recombine to 
form a virtual photon is $(\psi_{\gamma,f}^{*})_{\lambda}^{h\bar{h}}$. 

The elastic scattering of the $q\bar{q}$ pair with 
squared momentum transfer $\Delta^2 = -t$ is 
described by the elastic scattering amplitude,  
$A_{el}^{q\bar{q}}(x,r,\Delta)$. (The notation in this
section follows S. Munier {\it et al}~\cite{ref.MSM}.)
Here $A_{el}$ is defined 
such that
the elastic $q\bar{q}$ cross section is
\begin{equation}
   \frac{d\sigma_{q\bar{q}}}{dt} = 
   \frac{1}{16\pi} \left|A_{el}(x,r,\Delta) \right|^2.
\end{equation}
It is convenient to  work in coordinate space and define
the S-matrix element at a particular impact parameter $b$
\begin{equation}
S(b) = 1 +  \frac{1}{2} \int d^2\Delta \,e^{i{\bf b}\cdot{\bf \Delta}} \,
A_{el}^{q\bar{q}}(x,r, \Delta) \;.
\end{equation}
This corresponds to ones intuitive notion of impact 
parameter only when the dipole size is small 
compared to the size of the proton.
The total cross section for the $q\bar{q}$ pair is found
by taking the imaginary part
of $iA_{el}$  or
in terms of the S-matrix element
\begin{equation}
   \sigma_{q\bar{q}}(x, r) = \mbox{Im}\;iA_{el}^{q\bar{q}}(x,r,0) =  \int d^2b\, 2 (1 -  \mbox{Re}\; S(b)).
\end{equation}
This motivates the definition of the $q\bar{q}$ differential 
cross section 
\begin{equation}
\frac{d\sigma_{qq}}{d^2b} = 2\,(1 - \mbox{Re}\;S(b)).
\end{equation}

The total $\gamma^* p$  cross section for transverse and longitudinal
virtual photons is then given by the imaginary part of the forward
scattering amplitude 
\begin{equation}
\sigma^{\gamma^* p}_{L,T}(x,Q^2) = 
\int d^2r \int \frac{dz}{4 \pi} \, \sum_{f} \left ( \gamma| \gamma\right)_{\scriptscriptstyle L,T}^{f}
\, \sigma_{q\bar q }(x, r) \, ,
\label{eq:satur1}
\end{equation}
with
\begin{equation}
\sigma_{qq}(x,r) =\int{d^2b} \,\frac{d\sigma_{qq}}{d^2b}.
\label{eq:diptot}
\end{equation}
 $(\gamma|\gamma)_{\scriptscriptstyle L,T}^{f}$ 
denotes the probability for a 
polarized  virtual photon 
to fluctuate into a $q\bar{q}$ pair with flavor $f$
\begin{eqnarray}
\label{eq:olapgg}
    (\gamma|\gamma)_{L}^{f} &\equiv& \sum_{h\bar{h}} (\psi_{\gamma,f}^{*})_{0}^{h\bar{h}} (\psi_{\gamma,f})^{h\bar{h}}_{0}  \\
   (\gamma|\gamma)_{T}^{f}&\equiv& \frac{1}{2} \sum_{\stackrel{\scriptstyle h\bar{h}}{\lambda=\pm1}} (\psi_{\gamma,f}^{*})_\lambda^{h\bar{h}} 
      (\psi_{\gamma,f})_\lambda^{h\bar{h}}.  
\end{eqnarray}

In the dipole picture elastic 
diffractive vector meson production appears 
in a similarly transparent way \cite{BrodskyVM}. 
We denote amplitude for a  vector meson 
to fluctuate into a $q\bar{q}$ pair by  
$(\psi_V)_\lambda^{h\bar{h}}$.  
This vector-meson wave function will be discussed below. 
The amplitude for 
elastic diffractive vector meson production  
with squared momentum transfer $\Delta^2 = -t$ 
and transverse (T) or longitudinal polarization (L) is 
\begin{equation}
    A_{L,T}(\Delta) = 
  \int d^2r \int \frac{dz}{4\,\pi} \int d^2b\, 
  (V|\gamma)_{\scriptscriptstyle L,T}\, e^{-i {\bf b}\cdot{\bf \Delta}}\, 
  2\,(1 -S(b))\,
\label{eq:avm}.
\end{equation}
Assuming that the S-matrix element is predominantly real 
we may substitute $2(1-S(b))$ with
$\frac{d^2 \sigma_{q\bar{q}}}{d^2b}$. 
Then, the elastic diffractive cross section is 
\begin{equation}
  \frac{d\sigma_{L,T}}{dt} = \frac{1}{16\pi} \left|
  \int d^2 {\bf r} \int \frac{dz}{4\,\pi} \int d^2b\, 
  (V|\gamma)_{\scriptscriptstyle L,T}\, e^{-i {\bf b}\cdot{\bf \Delta}}\, 
  \frac{d\sigma_{qq}}{d^2b}\, \right|^2
\label{eq:saturVM}.
\end{equation}
Note, it is the total dipole cross section at a particular impact parameter
which appears in this formula.  
S. Munier {\it et al}~\cite{ref.MSM} extracted $S(b)$ from 
the diffractive data using the formalism outlined above.

\section{Model Description}

In order to describe the data it is necessary 
to make specific assumptions about the dipole cross section and the
photon and vector meson wave-functions. 
First, we give a simple model for the dipole cross section  
and contrast this model with 
other models that have appeared previously. Then we define the gluon
structure function $xg(x,\mu^2)$ and the proton shape function $T(b)$
which enters the dipole cross-section. 
Then we define and discuss the properties of  the
vector meson wave-functions in some detail.   
Finally after the dipole cross-section and the wave-functions are
specified, we determine the model parameters  by fitting the
inclusive total cross section and elastic diffractive vector  meson data.

\subsection{The Dipole Cross Section}

The total cross section for  a small $q\bar{q}$ dipole  to
pass through a dilute gluon cloud is 
proportional to the dipole area, the strong coupling constant, 
and the number of gluons in the cloud~\cite{ref.FRS}
\begin{eqnarray}
\sigma_{q\bar{q}} = \frac{\pi^2}{N_c}r^2\alpha_s(\mu^2)
                                      xg(x,\mu^2) \;,
\label{eq:dip}
\end{eqnarray}
where $x\,g(x,\mu^2)$ is  gluon density
at some scale $\mu^2$. 
Now imagine that the density of gluons in the 
target is not small. Divide  the target 
into thin slices of thickness $dz$
\footnote{$z$ denotes the longitudinal direction
and not the quark light cone momentum fraction in this
paragraph.}.
 The probability that
a dipole at impact parameter $b$ 
does not suffer an inelastic interaction passing through
  one  slice of the proton is
\begin{equation}
P(b) =  1 - \frac{\pi^2}{N_c}r^2\alpha_s(\mu^2)
                                      xg(x,\mu^2) \rho(b,z) dz \;.
\end{equation}
Here $\rho(b,z)$  denotes density of the 
gluons within a proton and is normalized to one
\begin{eqnarray}
 \int d^2b\,dz\, \rho(b,z) = 1 \;.
\label{eq:normT}
\end{eqnarray}
Exponentiating this result, the probability that the
dipole suffers no inelastic interactions passing 
through the entire proton is $|S(b)|^2$
\begin{eqnarray}
  |S(b)|^2 &=& \exp\left(- \frac{\pi^2}{N_c}r^2\alpha_s(\mu^2)
                                      xg(x,\mu^2) T(b)\right)   \; .
\label{eq:noint}
\end{eqnarray}
Here $T(b)$ is the thickness function, 
\begin{equation}
T(b)= \int_{-\infty}^{\infty} dz\,
\rho(b,z) \;. 
\end{equation}
To obtain the total cross section at a given impact parameter 
we assume that the S-matrix element is predominantly real.  Then
the cross section at a given impact parameter $b$ 
is given by $2 (1 - \mbox{Re}S(b))$ or
\begin{equation}
\frac{d\sigma_{q\bar{q}}}{d^2b} = 2\,\left[1-
\exp\left(-\frac{\pi^2}{2\,N_C}r^2\alpha_s(\mu^2)xg(x,\mu^2)T(b)\right)
\right]  \; .                 
\label{eq:dipgm}
\end{equation}

\begin{figure}[tbhp]
\includegraphics[height=10.cm,width=9.5cm]{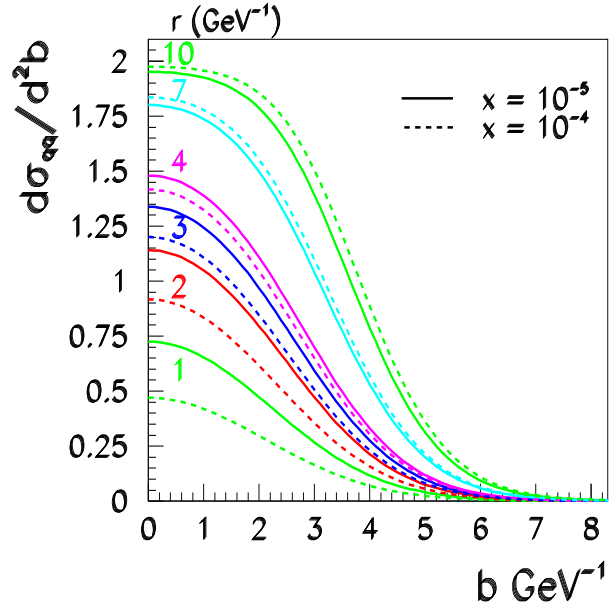}
\caption{\label{fig:sighb} The dipole cross-section  
as a function of the impact
  parameter $b$ for various dipole sizes  evaluated at two $x$-values.}
\end{figure}

This expression is known as the Glauber-Mueller dipole cross
section~\cite{ref.GlMu} and can also be obtained within the 
McLerran-Venugopalan model~\cite{ref.McLV}. It provides 
a simple model for the interaction of a $q\bar{q}$ probe with
a dense target. The function $T(b)$ and $xg(x,\mu^2)$ are determined from
the fits to the data. Figure~\ref{fig:sighb} shows the dipole cross
section as a function of the impact parameter $b$ for various dipole
sizes  at two distinct $x$-values. The properties of the dipole
cross-section will be discussed below. Similar dipole cross sections
were found in  the model of Ref.~\cite{Steffen}.
The cross-section is evaluated for the largest dipole sizes only to
illustrate saturation in the 
Glauber-Mueller approach. This dipole size is not contributing to the
observed cross-section. 

The Glauber-Mueller dipole cross section is a simple model
which  respects unitarity. As discussed below, in the large $Q^2$ region the model  
matches smoothly onto DGLAP since
we will evolve  $x g(x,\mu^2)$ with DGLAP. In 
the small $x$  region, 
the model evolution  can be improved. Indeed, evolution 
based upon the Balitsky-Kochegov 
equation \cite{BALITSKY,KOV} has been extensively 
studied \cite{LevinTotal,LevinJ} 
and provides a reasonable description of the total 
and total diffractive  cross sections. 
Nevertheless, the Glauber-Mueller dipole cross 
section provides a simple model which exhibits 
saturation as a function of impact parameter in a reasonably generic way and
is therefore a useful tool in exposing saturation effects in the
data.

\subsection{The Proton Shape $T(b)$}

The impact parameter dependence of the dipole cross-section determines
the properties of the $t-$distributions in elastic diffractive 
processes.
Here the variable $t$ denotes
the squared four momentum transfer between the incoming and outgoing proton 
in the  diffractive process, $\gamma^*p\rightarrow Vp$.  
The $t-$distribution 
is directly related to the Fourier transform  of $T(b)$ via 
Eqs.\,\ref{eq:dipgm} and \ref{eq:saturVM}.
Presently, the best data on diffractive $t$-distributions are available
for vector-meson production. HERA experiments
reported~\cite{ref.HERAVM,ref.H1JP,ref.ZEUSJP} that all diffractive
vector-meson 
$t$-distributions can be described by 
a single exponent $B$ for $|t|<1$ GeV$^2$   
\be
\label{eq:expdif}
\frac{d\sigma^{\gamma^* p}_{VM}}{dt} \propto \exp(-B|t|) \;.
\ee
\begin{figure}[htp]
\includegraphics[height=11.5cm,width=8.5cm]{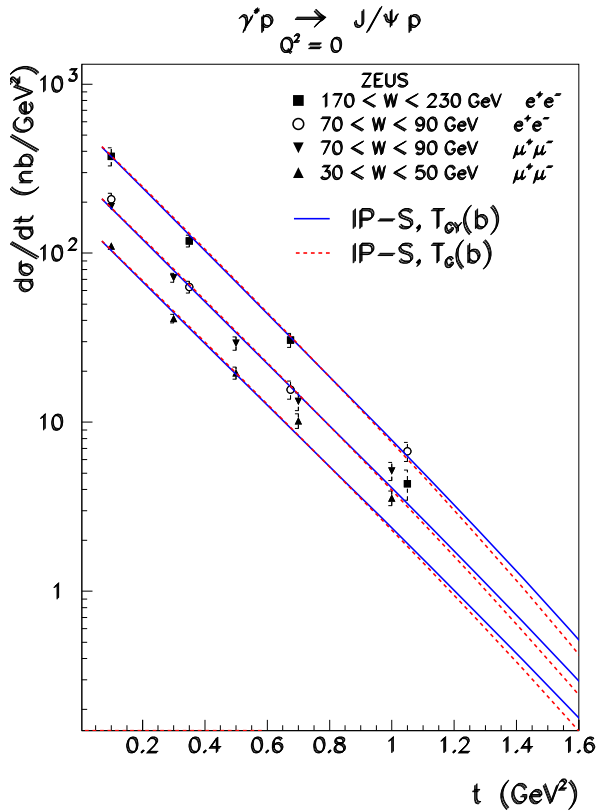}
\caption{ \label{fig.dst} The differential cross-section for exclusive
  diffractive $J/\psi$  production as a function of $t$ for
  representative bins in $W$~\cite{ref.ZEUSJP}. The solid (dashed) lines
  show  the results of the IP saturation model assuming  $T_{GY}$ ($T_{G}$) for the proton shape.}
\end{figure}

Since the $J/\psi$ wave-function is
better known than the $\rho$ or $\phi$ wave-functions, we will
concentrate on  $J/\psi$ production.  
The value of the coefficient $B$ for the
$J/\psi$ $t$-distribution  is around
4 GeV$^{-2}$ and is only weakly dependent 
$W^2$ and $Q^2$. The exponential form in Eq.~\ref{eq:expdif}
is the Fourier transform of  
\begin{equation}
\label{eq:Tb}
T_G(b) = \frac{1}{2\pi B_{G}} \exp(-b^2/2B_{G}) \;.
\end{equation}
We therefore tried this form for $T(b)$.
To account for the expected exponential behavior at
large impact parameters 
 we also considered the form
\begin{equation}
\label{eq:TbY}
T_{GY}(b) \propto \int d^2\, b' \exp(-({\bf b}-{\bf b'})^2/2B_{GY}) K_0(b'/W_{GY}) \;.
\end{equation}
Due to the asymptotic behavior of $K_0(x) \sim
\sqrt{\pi/2x}\exp(-x) $, $T_{GY(}b)$ falls off exponentially
at large impact parameters.

The  parameters $B_G$, $B_{GY}$ and $W_{GY}$ 
are only weakly dependent on the other parameters in the model and
were determined iteratively.
From the fit to the limited $t$ range, $|t| < 0.7$ GeV$^2$,
the following values were found: $B_G=4.25$ GeV$^{-2}$ for $T_G(b)$ and
$B_{GY}=3.25$ GeV$^{-2}$ and $W_{GY}=0.66$ GeV$^{-1}$ for $T_{GY}(b)$.
Figure~\ref{fig.dst} shows a representative subset of
data~\cite{ref.ZEUSJP} together with the results of a fit to the data
using both forms of $T(b)$.
The slope of the data fixes the parameters $B_{G}$ and $B_{GY}$ and $W_{GY}$.
The normalization of the curves in Figure~\ref{fig.dst} is a determined
by the underlying $J/\psi$ light cone wave function and is therefore
model dependent. We will use three model wavefunctions which
are discussed more fully in Sect.~\ref{vmwave} and in the Appendix. 
The figure is for the CORNELL wavefunction.
Unless explicitly stated, the profile function adopted throughout 
the rest of this work is $T_{GY}(b)$ which for practical purposes is equivalent
to $T_{G}(b)$.

\subsection{Gluon Structure Function and Dipole Evolution}
The dipole cross-section (Eq.~\ref{eq:dipgm})  requires 
the  gluon density $xg(x,\mu^2)$ for all scales $\mu^2$.   
As in  previous work~\cite{ref.BGK}, the scale $\mu^2$ is related to
the dipole size by
\be
\label{eq:scale}
\mu^2\,=\,\frac{C}{r^2}\,+\,\mu_0^2\;.
\ee
The gluon density  is evolved to the scale $\mu^2$
with leading order DGLAP without quarks for the gluon density. 
The evolution of the gluon density is performed numerically for every
dipole size $r$ during the integration of Eq.~\ref{eq:satur1}. 
The initial gluon distribution is taken at the scale 
$\mu_0^2$ in the form
\be
\label{eq:gluon}
xg(x,\mu_0^2)\,=\,A_g\,x^{-\lambda_g}\,(1-x)^{5.6}\, .
\ee
The parameters of the model  $C$, $\mu_0^2$, $A_g$, and $\lambda_g$
are determined from a fit to the DIS data.  
For light quarks the gluon distribution is 
evaluated at $x = x_{BJ} = \frac{Q^2}{Q^2 + W^2}$.  For
the charm quark the gluon structure function is evaluated
at $x = (4 m_c^2 + Q^2) / (Q^2+W^2)$, where $m_c$ is the charm 
quark mass~\cite{Kretzer}. This implies that 
the dipole cross section is flavor dependent when performing
the sum over flavors in Eq.~\ref{eq:satur1}.


The present model has, in some sense, one less parameter than the model of 
Ref.~\cite{ref.BGK}. The parameter $\sigma_0$  which 
previously determined the asymptotic size of the dipole
cross-section is now replaced by 
the parameter $B_G$ (or $B_{GY}$ and $W_{GY}$)  
which is (are) fixed by the shape of the diffractive $t$-distribution.

\begin{figure}[tp]
\includegraphics[height=9.5cm,width=9.5cm]{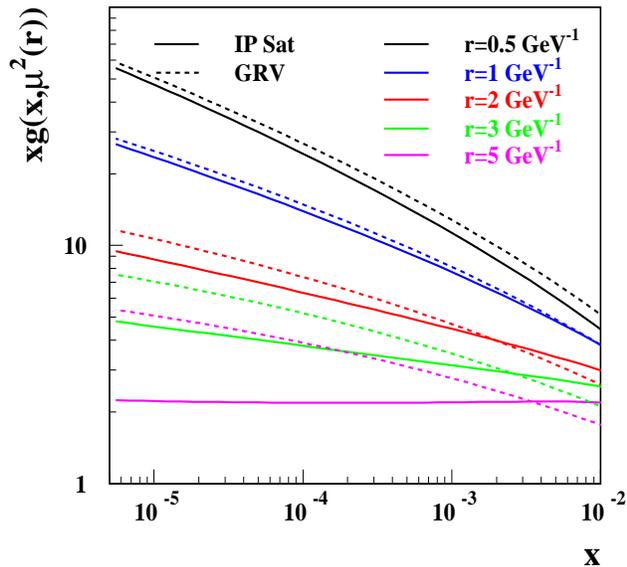}
\caption{\label{fig.xglur} The gluon structure function for various
  dipole sizes. The dipole size determines  the evolution scale
  $\mu^2$. The dashed lines shows the GRV results~\cite{GRV}.}   
\end{figure}
\begin{figure}[tp]
\includegraphics[height=9.5cm,width=9.5cm]{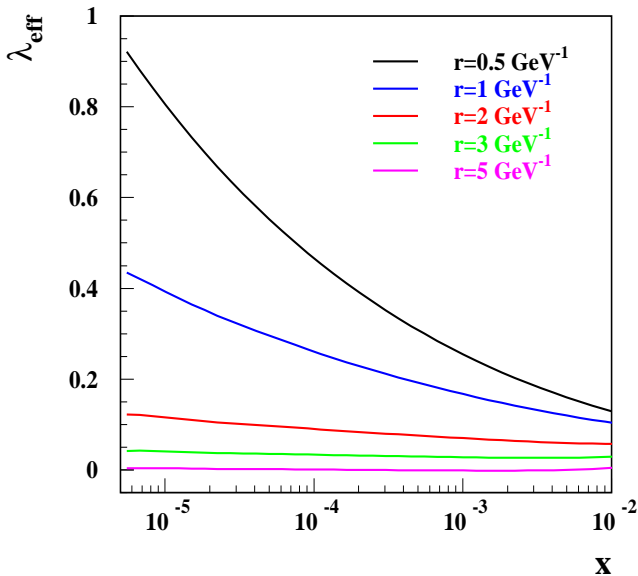}
\caption{\label{fig.lameff} The logarithmic rate of rise of the gluon
  structure function as a function $x$,
  $\lambda_{eff} = \frac{d \log(xg(x,\mu^2))}{d\log(1/x)}$.
  $\lambda_{eff}$ is a measure of the strength of the gluon emission
  process.}  
\end{figure}

The gluon structure function obtained from the fit is
shown in Fig.~\ref{fig.xglur} for various dipole
sizes.  The dipole size determines  the evolution scale $\mu^2$
through Eq.~\ref{eq:scale}.  QCD
evolution  changes  the rate of
rise of the gluon structure function $xg(x,\mu^2)$.  The rate of rise
is usually quantified by the exponent $\lambda_{eff}$, 
\be
\label{eq:lameff}
\lambda_{eff} = \frac{d \log(xg(x,\mu^2))}{d\log(1/x)}.
\ee
The exponent $\lambda_{eff}$
is a measure of the strength of  gluon emission process, see
e.g.~\cite{ref.BK}. 
Figure~\ref{fig.lameff} shows a rapid increase of the exponent
$\lambda_{eff}$ with decreasing $x$ for small values of the dipole
radius $r$.

\subsection{Comparison with the GBW model}

\begin{figure}[tp]
\includegraphics[height=8.5cm,width=8.cm]{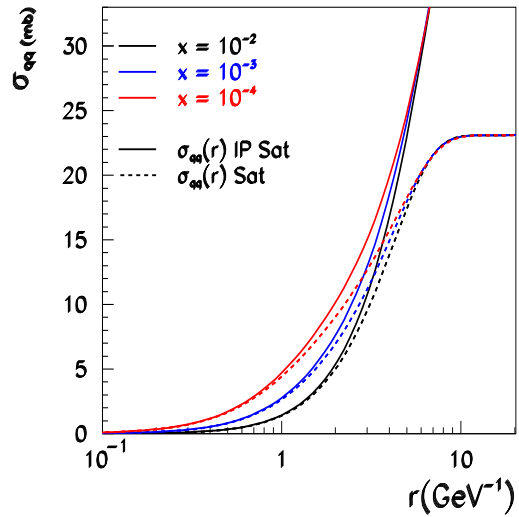}
\caption{\label{fig:dip.ori}The integrated dipole cross
  section as a function of the dipole size. The solid and dashed curves 
  trace the dipole cross-section in the IP Saturation model and the
  saturation model of Ref.~\cite{ref.BGK} respectively.}
\end{figure}

In the original GBW model~\cite{ref.GBW} 
the dipole cross section
$\sigma_{q\bar{q}}(x,r)$ was 
independent of impact parameter and gave a good description of
the inclusive total $\gamma^*p$ cross section. It 
also described some diffractive
processes~\cite{ref.GBW,ref.BGK,ref.CaldSo} without providing any
information about the $t$-dependence. 
In the GBW model the dipole cross-section 
$\sigma_{qq}$ is given by
\begin{equation}
\sigma_{q\bar{q}}^{GBW} = \sigma_0\,\left[1-
\exp\left(-\frac{r^2}{R_0^2}\right)\right] ,  
\label{eq:dipgbw}
\end{equation}
where $R_0$ denotes the saturation radius and $\sigma_0$ is a
constant. The saturation radius is $x$ dependent
\begin{equation}
 R_0^2 = \frac{1}{\rm GeV^2}
 \left(\frac{x}{x_0}\right)^{\lambda_{GBW}} .
\label{eq:rzero}
\end{equation}
The parameters $\sigma_0=23 $ mb, $\lambda_{GBW}=0.29$ and $x_0=3\cdot
10^{-4}$ were determined from a fit to the data. 
The saturation radius $R_0$ is analogous to the
gluon distribution and determines the growth of the   
total and diffractive cross sections with decreasing $x$.
For large size dipoles
$\sigma_{qq}^{GBW}$ saturates by approaching   a constant value $\sigma_0$ 
which is independent of the exponent $\lambda_{GBW}$.  

Once the impact parameter is included saturation becomes 
$b$ dependent. 
For small values of $b$, the cross-section grows rapidly with $r$
 until it  reaches the  saturation plateau seen in Fig.~\ref{fig:sighb},
$d\sigma_{qq}/d^2b=2$.
 The extension of the plateau $b_S$ can be
 estimated from the condition
 $\frac{\pi^2}{2\,N_C}r^2\alpha_s(\mu^2)xg(x,\mu^2)T(b_S)=1$. 
Using the Gaussian form of $T(b)$   and approximating
$xg(x,\mu^2)\sim  (x_0/x)^{\lambda_{eff}}$ we obtain  
\begin{equation}
 b_S^2 =   2B_G \, \log \left(
 \frac{\pi}{4\,N_C B_G}r^2\alpha_s(\mu^2)(x_0/x)^{\lambda_{eff}}  \right) \; .
\label{eq:bs}
\end{equation}
Thus, the integrated dipole cross-section $\sigma_{qq}$ increases
logarithmically for large $r$ and small $x$  due to the growth of the
plateau region.      
The logarithmic growth of the cross-section at small $x$  
is determined by $\lambda_{eff}$  and $B_G$. 
This behavior should be compared to the GBW cross section which becomes 
a constant $\sigma_0$ that is independent of $\lambda_{GBW}$. The differences
are illustrated in Fig.~\ref{fig:dip.ori}.

\subsection{Photon wave-function}
The longitudinal and transverse photon wave-function in the
conventions of Ref.~\cite{ref.BL} are given by 
\begin{widetext}
\begin{eqnarray}
   (\psi_{\gamma,f})_{0}^{h\bar{h}}(r,z)&=&
    e_{f} \delta_{f\bar{f}} \sqrt{N_c}
   \left[ -2 z (1-z)\,Q \left(\delta_{h+\bar{h}-} + \delta_{h-\bar{h}+}\right) 
    \right] \frac{K_{0} (\epsilon r)}{2\pi} \nonumber  \\  %
   (\psi_{\gamma,f})_{+1}^{h\bar{h}}(r,z) &=& 
    e_{f} \delta_{f\bar{f}}
  \sqrt{2 N_c} \left[ 
  -ie^{+i\varphi_r}\left( z\delta_{h+\bar{h}-} - (1-z)\delta_{h-\bar{h}+} \right)
   \partial_{r}
 + m_f\delta_{h+\bar{h}+} \right] \frac{K_0(\epsilon r)} {2 \pi}\\  \nonumber 
   (\psi_{\gamma,f})_{-1}^{h\bar{h}}(r,z) &=& 
    e_{f} \delta_{f\bar{f}}
  \sqrt{2 N_c} \left[ 
  +ie^{-i\varphi_r}\left( (1-z)\delta_{h+\bar{h}-} -
  z\delta_{h-\bar{h}+}  \right)
   \partial_{r}
 + m_f\delta_{h-\bar{h}-} \right] \frac{K_0(\epsilon r)} {2 \pi}
\label{eq:phot-wave} 
\end{eqnarray} 
\end{widetext}
where $r e^{i\varphi_r}=r_x + i r_y$,\, $\epsilon^2 = z(1-z)Q^2 + m_f^2$, 
$e_f$ is the electromagnetic charge of quark flavor $f$, and 
$\partial_{r} K_{0}(\epsilon r) = -\epsilon K_{1}(\epsilon r)$.

For small dipoles ($\epsilon  r \ll 1$) 
the photon wave-function scales as $K_1(\epsilon r)\sim 1/\epsilon
r$. For  large dipoles ($\epsilon  r \gg 1$),  
the  wave-function falls off exponentially with  
 the quark mass,   $K_1(\epsilon r) \sim
\sqrt{\pi/2\epsilon r}\exp(-\epsilon r) $. 

\subsection{Vector Meson Wave-Function}
\label{vmwave}

To describe vector meson production, we developed several models
of the $J/\psi$ light cone wave-function, $(\psi_{V})_{\lambda}^{h\bar{h}}$.
These phenomenological wave-functions should satisfy several model 
independent constraints~\cite{FS,DKP}. 
First, the wave-function must satisfy
the normalization condition
\begin{eqnarray}
  1 &=& \int \frac{d^2k dz}{16 \pi^3} \sum_{h\bar{h}} 
  \left|(\tilde{\psi}_V)^{h\bar{h}}_\lambda(k,z) \right|^2  .
\end{eqnarray}
Second, the decay width of the vector meson further constrains
the wave-function.  
Precise definitions
and explicit formulas for the decay width and other constraints on the
longitudinal and transverse wave functions  are given in the Appendix.   

As an example, we model the $J/\psi$ wave-function as a  Gaussian 
in momentum space in the meson rest frame
\begin{equation}
\tilde{\psi}_{NR}(p) = N_{L,T} \exp(- \frac{p^2}{\omega^2_{L,T}}) \;,
\end{equation}
were $\vec{p}$ denotes the charmed quark 3-momentum.
In DIS reactions the meson wave-functions
are defined on the light-cone and not in the $J/\psi$ rest
frame.  
Following Frankfurt and Strikman~\cite{FS} 
we boost the $J/\psi$ wave-function from 
the rest frame to the light-cone. The constants $N_{L,T}$ and $\omega_{L,T}$   were adjusted 
to satisfy the normalization and decay constraints.
This wave-function is denoted Gaus-RF (Rest Frame). 

As noted in Ref.~\cite{FS} this procedure does not lead to the
proper  asymptotic form of the light cone wave function at small
distances, $\tilde{\phi} \sim z(1-z)$. Therefore, we also considered  a
Gaussian wave-function defined directly on the light cone
\begin{equation}
\tilde{\phi}_{L,T}(k,z) = N_{L} \,z (1 -z)\, \exp(- \frac{k^2}{\omega^2_{L}}) \;.
\end{equation}
Once again, the norm and decay width constrain $N_L$
and $\omega_L$. This wave-function is denoted Gaus-LC (Light Cone).

 $J/\psi$  wave-functions were  also studied with  potential
models~\cite{Eichten,Buchmueller,EichQuig} and successfully describe 
the energy levels of $c\bar{c}$ bound states. 
Potential models were less successful at reproducing the meson decay
width which is sensitive to the  
short distance behavior of the wave-function. 
This shortcoming  may  be less
important since the short distance behavior of the wave-function is
suppressed by the vanishing dipole cross section at small $r$. 
We determined the non-relativistic
wave-function  by integrating 
the radial  Schr\"odinger equation with the Cornell potential~\cite{Eichten}.
As before we naively boosted this wave-function to the light-cone. 
Naturally, this wave-function  is denoted  CORNELL.

\begin{figure}[b]
\includegraphics[height=8.5cm,width=8.5cm]{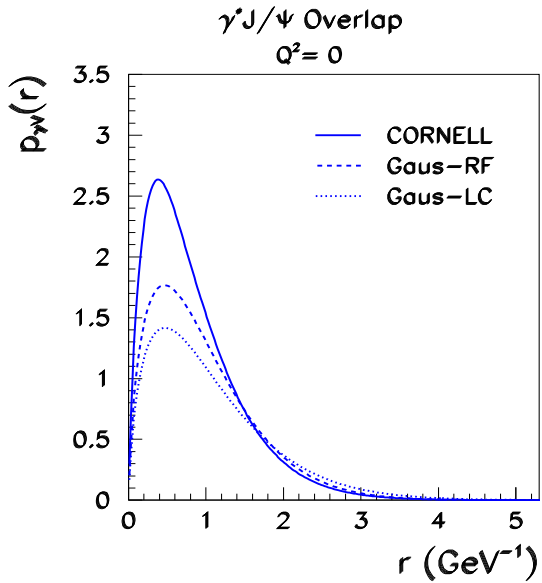}
\caption{\label{fig:psigaj}
  A comparison of the meson photon overlap function (Eq.~\ref{photmesolap}) 
  for the three different $J/\psi$ wave-functions adopted in this work. 
}
\end{figure}
In order to visualize the meson wave-function we define the meson
photon overlap $ p_{\gamma^{*}V}(r)$ as 
\begin{equation}
\label{photmesolap}
   p_{\gamma^{*}V} (r) = 
       r \int \frac{dz}{4\pi} (\gamma|V)_T \; ,
\end{equation}
which has the simple interpretation as the amplitude for 
diffractive vector meson production from the elastic scattering 
of a dipole of radius $r$.
The overlap function is  shown in
Fig.~\ref{fig:psigaj} for the transverse component of the three
$J/\psi$ wave-functions considered in the paper.

\subsection{Dipole size} 
\begin{figure}[t]
\hspace{-1.0cm}
\includegraphics[height=13.5cm,width=10.cm]{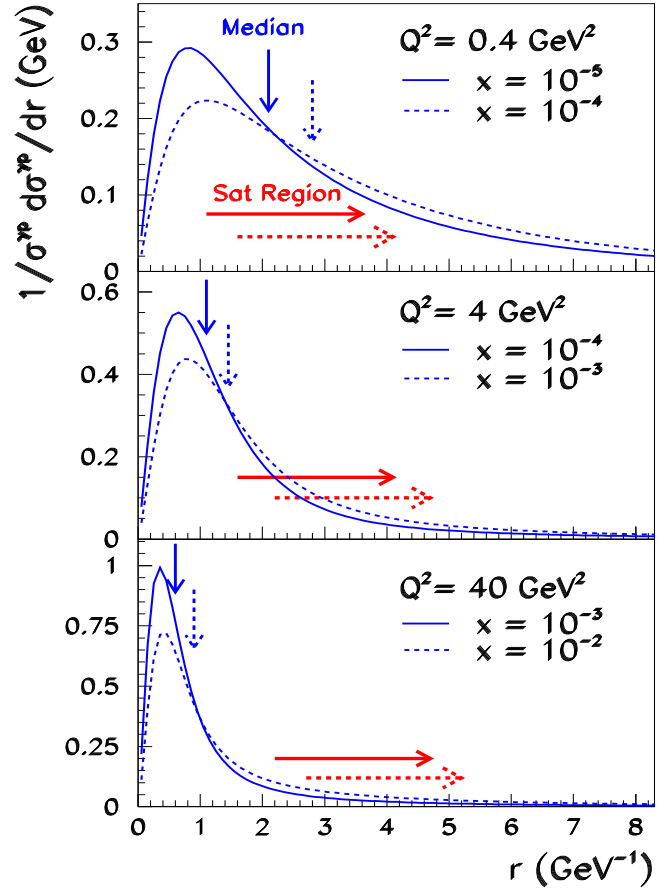}  
\caption{\label{fig.psitlr}
  The distribution of the dipole sizes contributing to the
inclusive total cross-section at various virtualities $Q^2$ and
some representative $x$ values. The vertical arrows indicate the
  median radius of a dipole which scatters off the proton. 
  The horizontal arrows indicate 
  saturation region  $r>r_{s}(b,x)$ at $b=0$. See Table~\ref{tab.sat}.}   
\end{figure}

To interpret the IP Saturation model physically,
we determined the typical
dipole size contributing to the total
cross-section. The
distribution of dipole sizes 
$ \frac{1}{\sigma^{\gamma^* p}}\frac{d\sigma^{\gamma^* p}}{dr}$
is the integrand of 
Eq.~\ref{eq:satur1}  after the $z$
integration is performed. Figure~\ref{fig.psitlr} shows this
distribution at representative values of $x$ and $Q^2$.
  For each illustrated $x$ and $Q^2$ the arrow indicates the median radius 
of a dipole which scatters off the proton. Note that at smallest $Q^2$ and
$x=10^{-5}$ the majority of dipoles are smaller than $r \alt 2$
GeV$^{-1}$ and thus half of the dipoles 
are perturbative.   

\section{Results for $\gamma^* p$ Processes}
\subsection{Inclusive total  cross-section}
\begin{figure}[t]
\includegraphics[height=12.5cm,width=9.5cm]{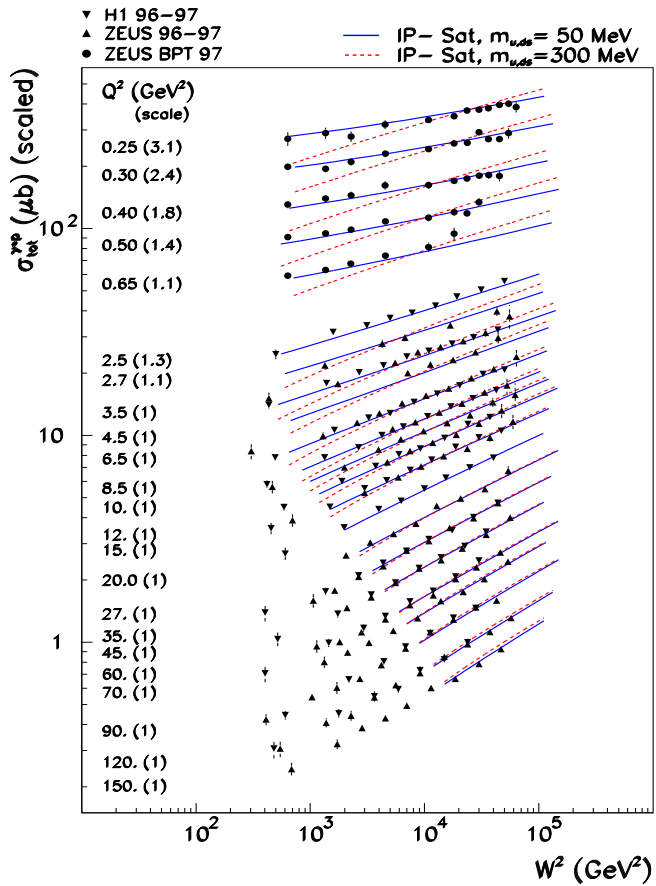}  
\caption{\label{fig:sigtotw2}
 The $\gamma^* p$ cross section as a function of $W^2$ at various
$Q^2$. The values of $Q^2$  are shown on the left side   of the 
plot along with the scale factor applied
to the data for a better visibility. The full (dashed) line shows a fit by the
IP Saturation Model with $m_{u,d,s}=50$ MeV ($m_{u,d,s}=300$ MeV).  The fits are performed for $x < 0.01$.
}
\end{figure}

The fit to the total $\gamma^*p$ cross-section data was performed 
in the range $x<0.01$ and $Q^2 > 0.25$ GeV$^2$. The small $x$
cut leads to an upper limit on
$Q^2$, \, $Q^2<200$ GeV$^2$.
Figure~\ref{fig:sigtotw2} shows the $\gamma^* p$ cross section as a
function of $W^2$ at the various $Q^2$  measured by the HERA
experiments~\cite{ref.sigtot}.
\begin{figure}[htp]
\includegraphics[height=8.5cm,width=8.5cm]{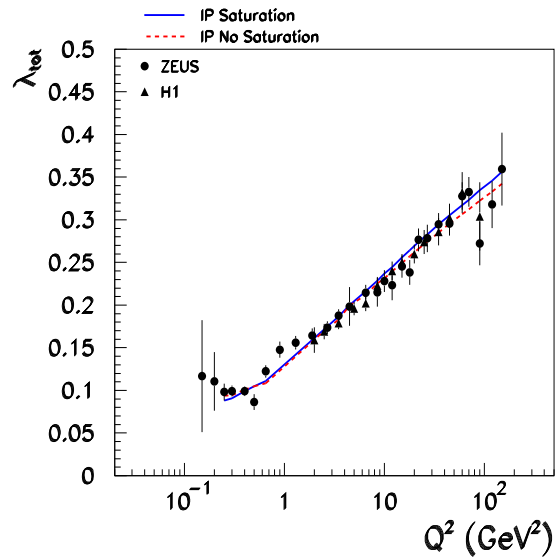}  
\caption{\label{fig.lam}
  The exponent $\lambda_{tot}$  as a function of $Q^2$. $\lambda_{tot}$ 
  parametrizes the total cross section 
  at fixed $Q^2$, $\sigma_{tot}^{\gamma^* p} \sim (1/x)^{\lambda_{tot}}$.  
  }
\end{figure}
Following  the rise of the total cross section with $W^2$ at 
fixed $Q^2$, 
we observe a  striking change: 
for small $Q^2$ (the top of Fig.~\ref{fig:sigtotw2})  
the cross section rises slowly with the center of mass energy $W^2$, 
while for large $Q^2$ (the bottom of Fig.~\ref{fig:sigtotw2})  
the cross section rises strongly. If we 
parameterize the cross section  by a power law 
\begin{equation}
\sigma_{tot}^{\gamma^*p}
\sim (W^2)^{\lambda_{tot}(Q^2)}\sim (1/x)^{\lambda_{tot}(Q^2)}\;, 
\end{equation}
this change in the rate of rise  from low to high $Q^2$ translates into a $Q^2$ 
dependence of the exponent $\lambda_{tot}$. $\lambda_{tot}(Q^2)$ is 
illustrated in Fig.~\ref{fig.lam} and summarizes one of the
principle observations at HERA.
At small $Q^2$, $\sigma_{\gamma^{*}p}$ increases with $W^2$ as in all 
hadronic reactions, $\lambda_{tot}  \approx 0.08$~\cite{ref.DonLan}. At 
at large $Q^2$ the exponent $\lambda_{tot}$ increases substantially.

In the IP Saturation model the increase of
the exponent $\lambda_{tot}$ with $Q^2$ 
is a consequence of the DGLAP evolution. 
When the dipole size becomes  smaller than $\sim \frac{2}{\mu_{o}}$
DGLAP evolution becomes  increasingly effective and
the growth of the gluon distribution at small $x$  becomes increasingly rapid. 
Mathematically speaking, $\lambda_{eff}$ increases for small size dipoles as 
seen in Fig.~\ref{fig.lameff}. 
Averaging over dipole sizes with the photon wave function 
translates  $\lambda_{eff}(r)$ into $\lambda_{tot}(Q^2)$. 
The large increase in $\lambda_{tot}$ with $Q^2$ indicates a
transition from a predominance of large to a predominance of small
dipoles when evaluating the cross-section. The contribution of large
dipoles at low $Q^2$ is determined by the light quark mass, $m_{u,d,s}$.

\begin{table} 
\hspace{1.6cm}
\begin{tabular} {|c|c|c|c|c|}  \hline
$m_{u,d,s}$  (MeV) & $\;\;\;\;\; A_g\;\;\;\;\;$  &  $\;\;\;\;\;\lambda_g\;\;\;\;\;$ & $\;\;\;\;\;\mu^2_0\;\;\;\;\;$ & $\;\;\;\;\;\chi^2/N_{DF}\;\;\;\;\;$ \\ \hline\hline
 300         &    1.67   &  0.226  & 6.85  & 702/160     \\ \hline
 150         &    2.34   &  0.043  & 1.32  & 199/160     \\ \hline
 100         &    2.85   &  -0.04  & 0.96  & 153/160     \\ \hline
 50          &    3.47   &  -0.12  & 0.77  & 137/160     \\ \hline
\end{tabular}
\caption{Table of parameters determined in the fit to the total
  $\gamma^* p$ cross-section for different assumptions on the light
  quark mass $m_{u,d,s}$.}
\label{tab.fit}
\end{table}

The parameters $A_g$, $\lambda_g$ and $\mu_0^2$ were found by fitting the
total $\gamma^* p$ cross-section for various  
light quark masses. The parameter $C$ is strongly
correlated to the parameters $A_g$ and $\lambda_g$ 
and does not affect the fit quality.
Therefore, we set $C=4$   by
fiat.  The mass of the charm quark was 
$m_{c}=1.3$ GeV as required by elastic diffractive $J/\psi$ 
production discussed below. The results of the fits to the ZEUS data
are collected in 
Table~\ref{tab.fit}.   
 Because of the good agreement between the H1
and ZEUS data sets the combined H1-ZEUS fit returns nearly identical
parameters. 
The fits show a clear preference for the light quark
mass of  $m_{u,d,s}=50$ MeV, which is taken as a
reference value. 
Note, the quality of the fit reported here is even
better than the fits of Ref.~\cite{ref.BGK}.  


As seen in Table~\ref{tab.fit}, when light 
quark mass is increased, the fit subsequently increases $\lambda_g$. 
DGLAP evolution can only increase the observed rate of rise of the
cross section with $W^2$,  $\lambda_{tot}>\lambda_g$. Therefore, as seen 
in Fig.~\ref{fig:sigtotw2},  for
larger masses of the light quarks the cross section rises
too quickly with $W^2$ at low $Q^2$.   
It is certainly known that the strange quark mass is heavier than
the $u$ and $d$ quarks. However, the influence of the strange quark
on the overall fit is relatively small since its contribution to 
$F_{2}$ is suppressed relative to the $u,d$ contribution by 
the  ratio of charges squared, $e^2_{s}/(e^2_{u} + e^2_{d}) = 1/5$. Therefore,
we have neglected the difference between the $s$ and $u,d$ quark masses. 
Including a heavier strange quark would make the fit only 
marginally worse. However,
the clear dependence of the fit quality on the mass of the light
quark is a property of the model. It is due to the interplay of 
DGLAP evolution and the properties of the photon wave function.
It may indicate the shortcomings of the wave function at large distances.
 
In contrast to the GBW model~\cite{ref.GBW} and the
model of Ref.~\cite{ref.BGK}, saturation
is not responsible for the change in $\lambda_{tot}$ 
with $Q^2$ once the impact parameter is included. 
This is because  the total dipole cross section
continues to grow within the saturation region 
although the differential dipole cross-section has attained its 
asymptotic value $\frac{d\sigma}{d^2b}=2$.
Therefore, for the total cross-section saturation effects are smaller
than in the  original GBW model.  However,
 for diffractive $t$-distributions saturation will appear more clearly.  

\subsection{Charm Production}

\begin{figure}[htp]
\includegraphics[height=10.5cm,width=9.5cm]{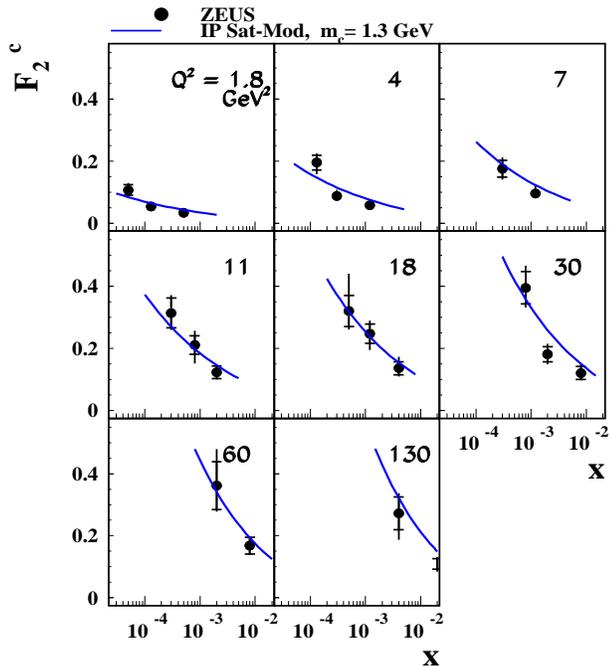}
\caption{\label{fig:f2ch}
  A comparison of the measured $F_2^c$ ~\cite{ref.ZEUSCH}
  to the results of the model.   
}
\end{figure}

The dipole model makes a direct prediction for the inclusive charm
contribution to 
$F_2$. In the dipole approach 
the charm quark distribution is calculated from the gluon
distribution.    Figure~\ref{fig:f2ch} shows
a comparison between  the measured and predicted values of the charm
structure function $F_2^c$.  The
 results depend weakly on the charm mass.    
The good agreement with data for both ZEUS and H1
experiments~\cite{ref.ZEUSCH,ref.H1CH} 
confirms the consistency of the  model and  supports 
the dipole picture.        
 
\subsection{Determination of $F_L$}
\begin{figure}[htp]
\includegraphics[height=9.5cm,width=9.5cm]{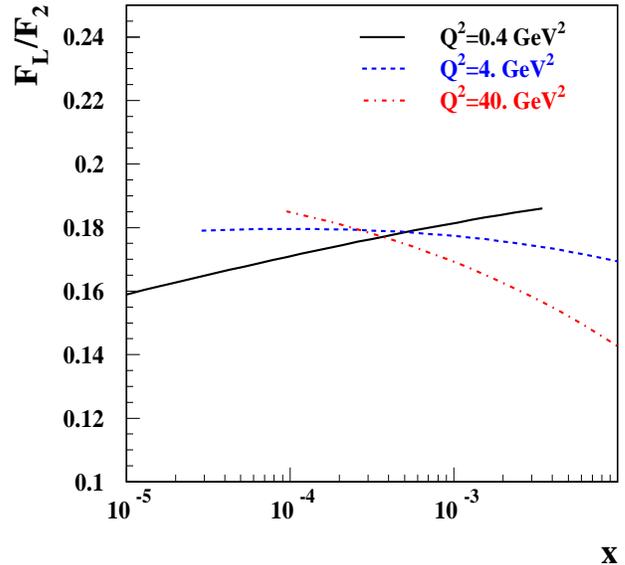}
\caption{\label{fig.flfit}
  The ratio  $F_L/F_2$ as a function of $x$ at various
photon virtualities.
 }  
\end{figure}
\begin{figure}[htp]
\includegraphics[height=9.cm,width=9.cm]{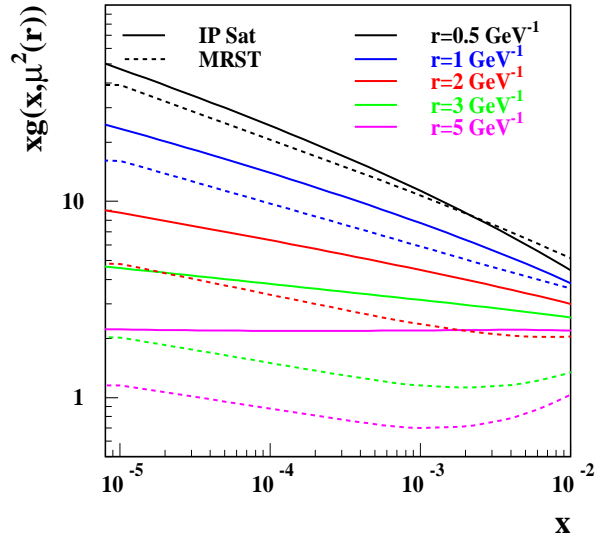}
\caption{\label{fig.xglum}
 A comparison of the gluon structure function of the
  IP Saturation  model with the results of the MRST fit~\cite{MRST} for
  various dipole sizes. The
  dipole size determines the evolution scale $\mu^2$. 
 }  
\end{figure}

As a byproduct of our investigation  we determined the
longitudinal structure function $F_L$. A measurement of $F_L$  would be
 a critical test of our understanding of the gluon distribution.
 $F_L$ is zero in the naive parton model and is non-zero only after
gluon radiation is included.
 In dipole models at small $x$, both $F_L$ and $F_2$ are governed  by the gluon distribution
and therefore they behave similarly.
Figure~\ref{fig.flfit} shows the ratio of $F_L$ to $F_2$ as a function of
$x$ for various photon virtualities. The ratio  is around 18 $\%$ and
 gently varies with $x$.   
In the standard QCD fits $F_L/F_2$ may show a 
stronger variation with $x$   at low $Q^2$ \cite{RThorn}. This
is because 
the shape of the gluon structure function 
seen in Fig.~\ref{fig.xglum} is  very
 different  from the dipole model presented here. 
The difference in the gluon structure
function between the MRST fit~\cite{MRST} and the dipole model fit is due to 
the sea quark contribution.  In the MRST fit the initial sea quark
contribution is independent of the gluon distribution and subsequently  
influences the gluon evolution.  In the dipole fit the sea quark distribution is
always derived from the gluon distribution.  It is important to
measure $F_L$ in order to determine the gluon density precisely and to
clarify the relative roles of gluons and sea quarks.

\subsection{$J/\psi$ production} 

\begin{figure}[htp]
\includegraphics[height=8.5cm,width=8.5cm]{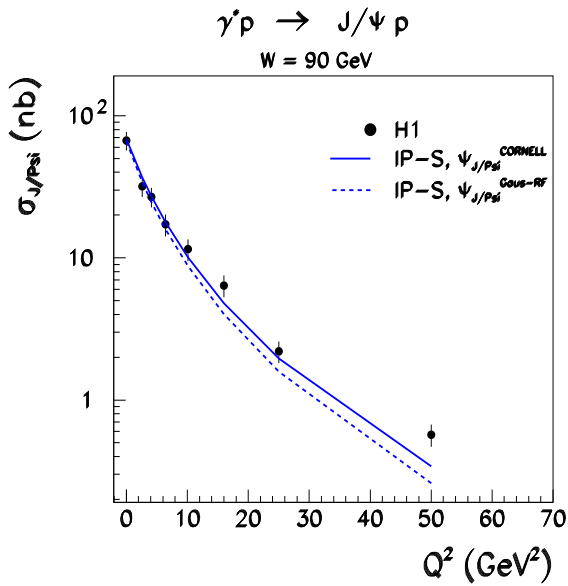}
\caption{\label{fig:totq2}
  A comparison of the measured total elastic diffractive 
  $J/\psi$ cross-sections~\cite{ref.H1JP} with
  the results of the model.   
}
\end{figure}
\begin{figure}[htp]
\includegraphics[height=8.5cm,width=8.5cm]{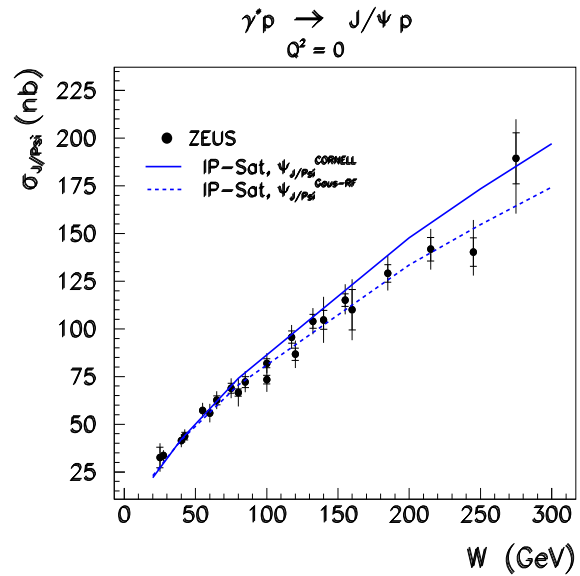}
\caption{\label{fig:jp}
  A comparison of the measured total elastic diffractive $J/\psi$
  cross-sections~\cite{ref.ZEUSJP} with 
  the results of the model.   
}
\end{figure}

The cross sections for elastic diffractive  $J/\psi$ production were calculated using
Eq.~\ref{eq:saturVM} with the dipole cross section determined by 
fitting $\sigma_{\gamma^{*}p}$.  Figure~\ref{fig:totq2} compares  the
computed  
total $J/\psi$ cross-section with  H1 measurements~\cite{ref.H1JP} 
for various photon virtualities  at $W=90$ GeV.  
The absolute magnitude of the
cross-sections is very sensitive to the  charm quark mass. 
The charm quark mass 
 was adjusted to $m_c=1.3$ GeV in order to match the  experimental
 cross section at $Q^2=0$. A small
 shift of 50 MeV in the charm mass  
changes the cross-section by around 15$\%$. 

The predicted total $J/\psi$ cross-sections for the
 Gaus-RF and Gaus-LC wave-functions are very similar in the 
  $Q^2$ range considered. Therefore, we only show
 the predictions of the Gaus-RF wave-function in what follows. The CORNELL
 wave-function, which is more 
 concentrated at the origin leads to a 
 prediction which is in significantly better
 agreement with data at large $Q^2$, see  Fig.~\ref{fig:totq2}. 
In contrast to Ref.~\cite{FS},  the 
scale $\mu^2 = C/r^2 + \mu_{0}^2$ is fixed with 
$C=4 $.
Within the range $C=1\sim 10$ the results were almost independent  of
the parameter $C$ after  refitting total $\gamma^{*}p$ cross-section.

Figure~\ref{fig:jp}  compares model predictions  with the
measured $J/\psi$ cross-section as a function 
of energy $W$  at $Q^2=0$.  In this case, both $J/\psi$ wave-functions
are in good agreement with the  data. 

Note, there is a   $ \sim 15\%$ discrepancy in the normalization of
the  $J/\psi$ cross-sections between ZEUS and H1.
 This discrepancy was absorbed in the
  charm quark mass; 
we take $m_c=1.25$ GeV when comparing with  ZEUS data
and $m_c=1.3$ GeV when comparing with H1.

\subsection{$t$ distributions for $J/\psi$ production} 
\begin{figure}[t]
\includegraphics[height=11.cm,width=8.5cm]{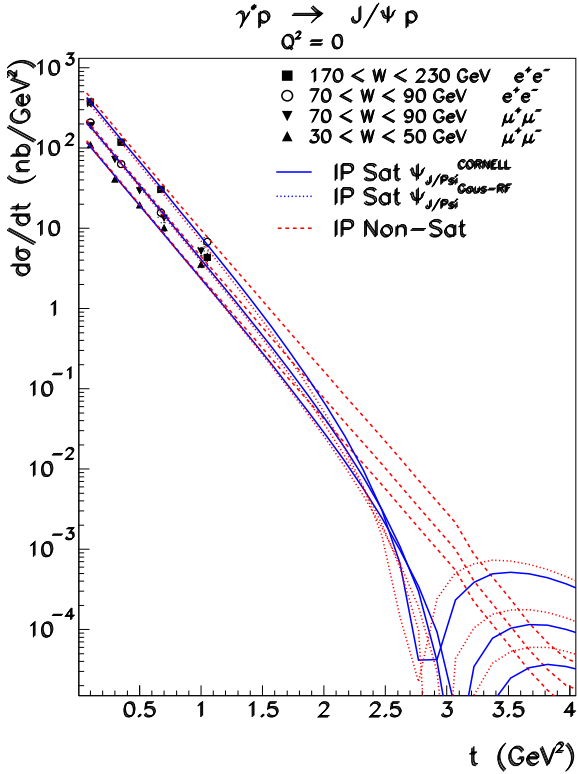}
\caption{\label{fig:dsl}
A  comparison of the measured diffractive $J/\psi$
cross-sections~\cite{ref.ZEUSJP} with 
   the results of the IP Saturation 
   model for two different $J/\psi$ wave-functions.
  Also shown are the results of  the non-saturated model 
  evaluated with  CORNELL wave-function.   
}
\end{figure}
\begin{figure}[htp]
\includegraphics[height=11.cm,width=8.5cm]{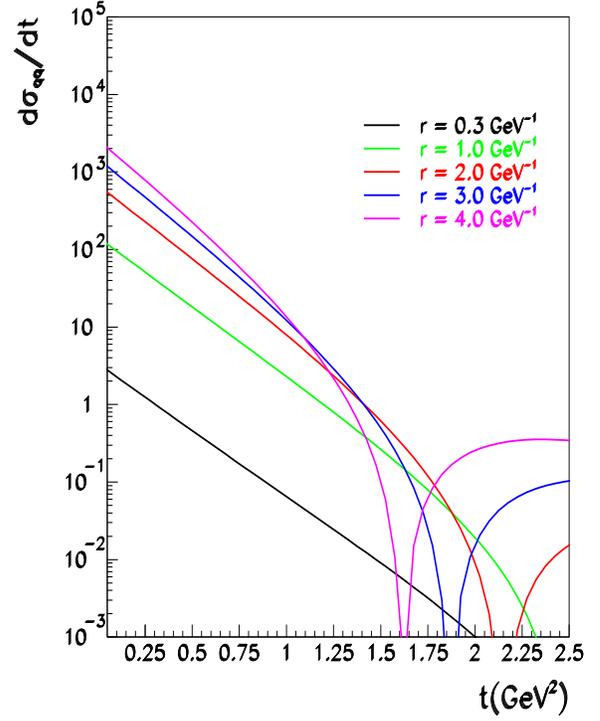}
\caption{\label{fig:ftshb}
  The Fourier transform of the differential dipole $\frac{d\sigma_{qq}}{d^2b}$
cross section for various dipole sizes.   
}
\end{figure}
Figure~\ref{fig:dsl}  compares  the  $J/\psi$
diffractive cross-section
as a
function of $t$ for various $W^2$ to model predictions. 
In the measured region, both 
$J/\psi$ wave-functions lead  to very similar
results although  at large $t$ they start to differ significantly. 
In the large $t$ region the exponential behavior of the
 differential cross section is considerably
 distorted and the IP Saturation model predicts diffractive dips.
Figure~\ref{fig:ftshb}  illustrates the emergence of 
the dip by displaying the Fourier transform of the differential dipole
cross section  for various dipole sizes 
\begin{equation}
  \frac{d\sigma_{qq}}{dt} =  \left|
   \int d^2b\, 
          e^{-i {\bf b\cdot\Delta}}\, 
  \frac{d\sigma_{qq}}{d^2b}\, \right|^2 \; .
\label{eq:sqqt}
\end{equation}
For small $r$, the $b$ dependence of the dipole cross section is
essentially Gaussian and its Fourier transform is  a pure
exponential in $t$.  For large $r$, due to saturation, the
shape of the dipole cross section becomes more box-like, as seen in
Fig.~\ref{fig:sighb}. The two dimensional Fourier transform of a box
is  proportional to the Bessel function $J_1$ which  leads to dips in
the $t$ distribution.   
Although the distortions to the proton shape seen in 
Fig. \ref{fig:sighb} are small when displayed in coordinate space,
they appear vividly after 
taking the Fourier transform and examining the $t$-distribution.
For dipoles with $r\ge 3$ GeV$^{-1}$ the
$t$-distribution is substantially changed.

\subsection{Saturation versus Non-Saturation} 

\begin{figure}[b]
\includegraphics[height=9cm,width=9cm]{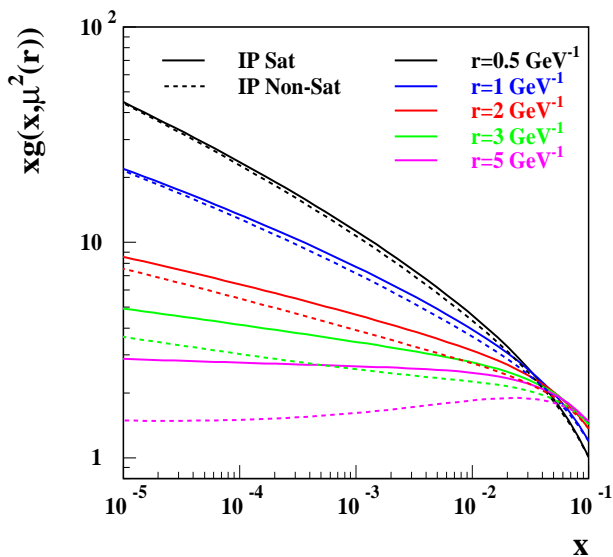}  
\caption{\label{fig:xgluns}
  A comparison of the gluon structure function in the
  IP Saturation  and IP Non-Saturation models at various dipole sizes. The
  dipole size determines the evolution scale $\mu^2$. 
 }  
\end{figure}

In order to isolate the effects of saturation we compared the results of the
IP Saturation model   with a model which is  not unitarized. 
This model, denoted IP Non-Sat, replaces the dipole cross-section of 
Eq.~\ref{eq:dip} with the first term after expanding of the exponential
\begin{equation}
\frac{d\sigma_{q\bar{q}}}{d^2b} = \frac{\pi^2}{N_C}r^2\alpha_s(\mu^2)xg(x,\mu^2)T(b)
 \; .                 
\label{eq:dipgmns}
\end{equation}
The proton shape $T(b)$ was assumed to be the same  as in the
saturated case. The 
fit to the total cross section was repeated and a non-saturated gluon
structure function was determined. 
The optimal fit was obtained for a higher quark mass than in the
saturated case, $m_{u,d,s}=150$ MeV. This is expected since the
dipole cross-section $\sigma_{q\bar{q}}$ grows faster 
with $r$ than in the saturated case.
The fit yields  $A_g=3.5$, $\lambda_g=-0.13$, $\mu^2_0=0.80$ and
$\chi^2/N_{DF}=0.8$. Naturally,  the saturated and non-saturated dipole
cross sections are  very similar since they both fit the inclusive $\gamma^* p$
cross-sections very well. Consequently, all observable cross-section
which only are sensitive to the dipole cross-section  integrated over 
impact parameter are indistinguishable. This was verified explicitly
for the total diffractive $J/\psi$ production.  In particular, this
agreement can be seen in Fig.~\ref{fig.lam} for  the
$\lambda_{tot}(Q^2)$ distribution. 
 However, the gluon distributions are not the same as illustrated in
Fig.~\ref{fig:xgluns} which compares
 the saturated and non-saturated models. 
For small dipole sizes the distributions
are  similar  but they differ significantly as $r$ becomes large.  

The $t$-distribution provides  additional
 insight into the dynamics 
  of  saturation  in DIS. Figure~\ref{fig:dsl} compares
 predictions for  diffractive $J/\psi$ production in the
saturated and non-saturated modes . As discussed above saturation
distorts the impact parameter distribution and  leads
unavoidably to diffractive dips in the $t$-distribution. 
Without saturation, as seen in Fig.~\ref{fig:dsl},  the exponential
decrease continues to large $t$.
  Diffractive dips in the $t$ distribution were clearly observed in 
elastic hadronic reactions~\cite{Goulianos}.  In  DIS processes at
HERA the measured range of the $t$-distribution is too small for a dip
to appear. 
In addition the HERA experiments have always measured  elastic
diffractive events with some small admixture of  inelastic diffractive 
events. Even if inelastic events are a small fraction of the
total, these events will
dominate the $t$-distribution at large $t$ since the
inelastic $t$-distribution  falls slowly~\cite{ref.ZEUSJP,Goulianos}.       

\begin{figure}[tp]
\includegraphics[height=9cm,width=9cm]{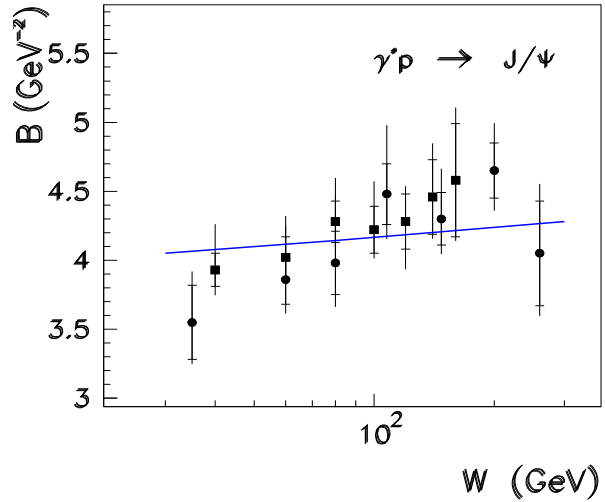}
\caption{\label{fig:bzslop}
  The diffractive slope $B$ measured by the ZEUS experiment 
 as a function of $W$ at $Q^2 = 0$. The solid line shows the prediction
  of the IP Saturation model. 
}
\end{figure}

In the small $t$ range, saturation slightly
increases the slope parameter $B$ of diffractive vector meson production
 at large $W$. 
Figure~\ref{fig:bzslop} compares
the slope parameter $B$  determined by the ZEUS
experiment with model predictions. The observed
 increase of the slope $B$ with $W$ (called shrinkage) 
 is slightly larger  than the model prediction. This could be an
 indication that  the shape $T(b)$ is somewhat 
dependent on $x$ due to QCD evolution.
Since
considerably more data is available than was used in the present analysis we
plan further study of the observed shrinkage.

Concluding we note that the $t$-distribution
provides  precise information about the interplay of saturation
and evolution phenomena. Therefore, it is  crucially important to measure 
elastic vector meson production with a small systematic uncertainty 
over a wide range in $t$.
\section{Saturation Scale $Q^2_S$}
\begin{figure}[t]
\includegraphics[height=9cm,width=9cm]{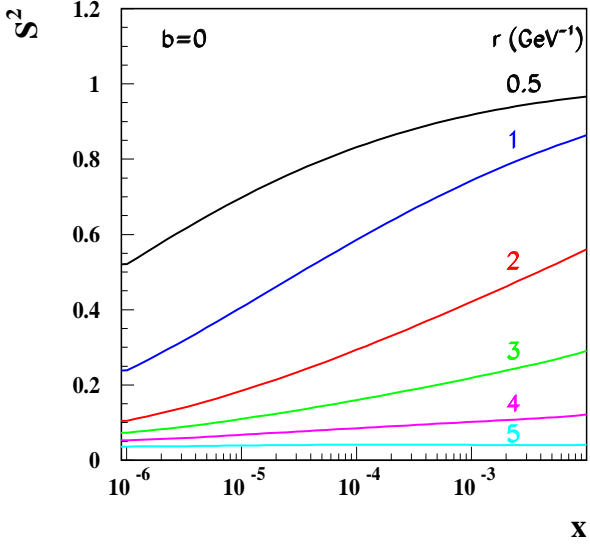}
\caption{\label{fig:s2mat}
 The square of the S-matrix as a function of $x$ for various dipole sizes.}
\end{figure}

In the limit of high gluon density called the Color
Glass Condensate (CGC)\cite{CGC}, the  interaction probability in
DIS becomes large. 
The probability of no interaction  is measured by the square of
the S-matrix element, $S^2(b)$.  
Normally, in high energy reactions the interaction probability is
small, $S^2\sim 1$. However, in the CGC  limit
this probability approaches unity,
$S^2\sim 0$. 
 Figure~\ref{fig:s2mat} shows  $S^2(b=0)$  as a 
 function of $x$ for various dipole sizes.  
For small size dipoles the gluon structure function grows strongly with decreasing $x$
and the S-matrix diminishes quickly. For large size dipoles,
the gluon density density does not grow and 
the S-matrix is small even at relatively large $x$. 

To delineate the saturation region in the ($x$,$r^2$) plane 
we first define  a density profile  $D$
\begin{equation}
D(b,x,r^2) = 
\frac{2\, \pi^2}{N_C}\alpha_s(\mu^2(r^2))xg(x,\mu^2(r^2))T(b) \; . 
\label{eq:q2subs}
\end{equation}
Then, the saturation radius $r_S$ is defined as the dipole size  for which
 proton consists of one interaction length 
\begin{equation}
S^2 = \exp\left(-\frac{D(b,x,r^2_S)\,r^2_S}{2}\right) = e^{-1} \; .
\end{equation}
This defines an implicit equation for $r^2_S$ which can be 
solved numerically. Now the saturation scale $Q^2_S$ is defined 
as the density profile $D$ evaluated at the saturation radius
\begin{equation}
Q^2_S(b,x) = D(b,x,r^2_S)= \frac{2}{r^2_S}.
\end{equation}
In the GBW model $S^2= \exp(-r^2/2R_0^2)$ and we find  $Q^2_S=1/R_0^2$ in
accord with the original definition of the saturation
scale. 
The saturation region is then defined  by the condition that $r>r_S(b,x)$.

\begin{figure}[tp]
\includegraphics[height=9cm,width=9cm]{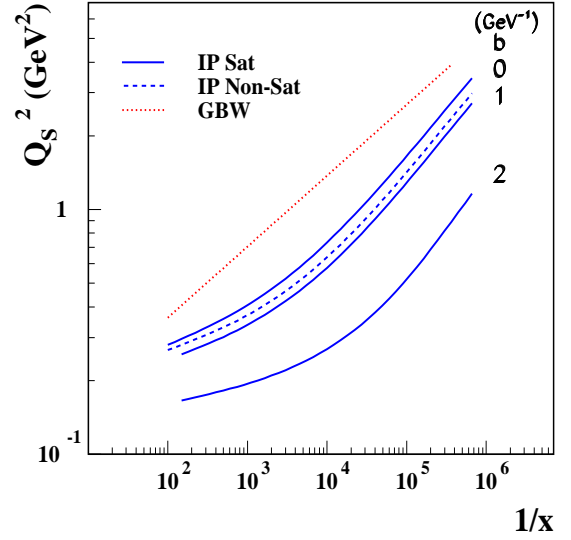}
\caption{\label{fig:q2sx}
  The saturation scale as a function of $x$. 
  The dashed curve shows the critical line ($S^2 = e^{-1}$)
  in the Non-Saturated model at $b=0$.
}
\end{figure}
\begin{figure}[htp]
\includegraphics[height=9cm,width=9cm]{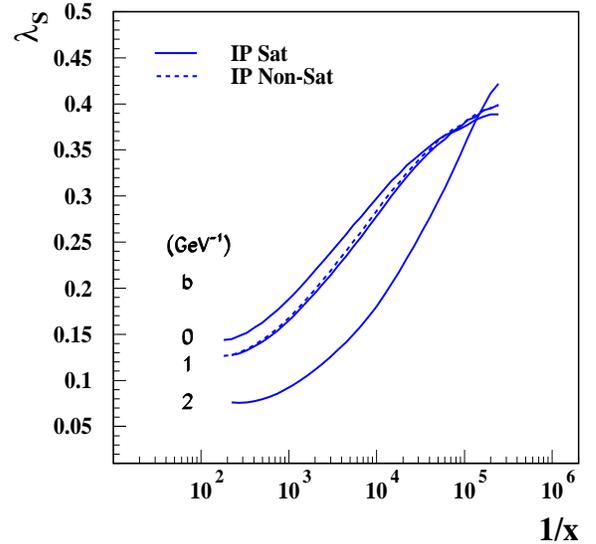}
\caption{\label{fig:lamsx}
  The saturation exponent $\lambda_S$ as a function of $x$ and impact 
  parameter. The dashed curve shows the exponent along the critical
  line ($S^2 = e^{-1}$) in the Non-Saturated model at b=0.
}
\end{figure}
Figure~\ref{fig:q2sx} shows the saturation scale $Q^2_S$ as a function
of $1/x$ in the central impact parameter region. 
We observe that the saturation scale  is lower  than the
GBW value. This is primarily because the charm quark contribution to
$F_2$  was neglected when the saturation scale was evaluated.
Including the charm quark reduces the gluon density and the
saturation scale.

We also calculated the S-Matrix element and determined the
 corresponding effective saturation scale for the non-saturated model.
As seen in  Fig.~\ref{fig:q2sx},
 the saturation scale is very similar for 
IP Saturated and IP Non-Saturated  models. This is
because the corresponding gluon structure functions are
very similar for small size dipoles, as seen in Fig.~\ref{fig:xgluns}. 
Therefore, our estimates of the saturation scale are only
weakly model dependent. 

Table~\ref{tab.sat} records the saturation
\begin{table}[b] 
\begin{tabular} {|c|c|c|c|}  \hline
$\;\;\;\;\; b{\rm \;\; (GeV^{-1})}\;\;\;\;\;$  &  $\;\;\;\;\; 0 \;\;\;\;\;$ &
 $\;\;\;\;\; 1\;\;\;\;\;$ & $\;\;\;\;\; 2\;\;\;\;\;$ \\ \hline
 $\;\;\;  Q^2{\rm \;\; (GeV^{2})}\;\;\; x\;\;\;$ & 
$\;\;\;r_S\;\;\;\; \%\;\;\;$ &
$\;\;\;r_S\;\;\;\; \%\;\;\;$ & $\;\;\;r_S\;\;\;\; \%\;\;\;$  \\ \hline\hline
0.4\;\;\;\;\; $10^{-5}$ & 1.1\;\;\; 72 & 1.3\;\;\; 68 & 2.0\;\;\; 52 \\ \hline 
0.4\;\;\;\;\; $10^{-4}$ & 1.6\;\;\; 70 & 1.9\;\;\; 65 & 2.7\;\;\; 51 \\ \hline
4.0\;\;\;\;\; $10^{-4}$ & 1.6\;\;\; 33 & 1.9\;\;\; 25 & 2.7\;\;\; 17 \\ \hline
4.0\;\;\;\;\; $10^{-3}$ & 2.2\;\;\; 44 & 2.4\;\;\; 28 & 3.2\;\;\; 20 \\ \hline
40.\;\;\;\;\; $10^{-3}$ & 2.2\;\;\; 13 & 2.4\;\;\; 12 & 3.2\;\;\;\; 9 \\ \hline
40.\;\;\;\;\; $10^{-2}$ & 2.7\;\;\; 21 & 2.9\;\;\; 18 & 3.5\;\;\; 16 \\ \hline
 
 \end{tabular}
\caption{Table of saturation radii and of the percentages of the
inclusive total $\gamma^* p$ cross-section in the saturation region, $r>r_S$.}
\label{tab.sat}
\end{table}
radii and the percentage of the  total $\gamma^*
p$ cross-section in the saturation region, $r>r_S$. At 
low $Q^2$ ($Q^2=0.4\,\mbox{GeV}^2)$ the cross-section is dominated by dipoles which 
are larger than the saturation radius.  As seen in Fig.~\ref{fig.psitlr}, a fraction of
these dipoles may be considered perturbative since they are
smaller than $r\alt2\,\mbox{GeV}^{-1}$.
In the large $Q^2$ region  saturation is of secondary importance. 
The region $Q^2\approx4$ GeV$^2$ is particularly
interesting since a significant fraction of the dipoles  are perturbative and 
are approximately the size of the saturation radius. 
BFKL and saturation dynamics are expected to influence observable cross
sections in this kinematic domain.

It is also interesting to define the saturation exponent $\lambda_S$ 
\be
\label{eq:lamsat}
\lambda_{S} = \frac{d \log(Q^2_S)}{d\log(1/x)}.
\ee
Figure~\ref{fig:lamsx} shows the saturation exponent $\lambda_S$ as a function
of $1/x$  for various impact parameters.
The saturation exponent exhibits a sizable $x$ dependence. It
varies between  $\lambda_{S}=  0.15$ at $x= 10^{-2}$ and
$\lambda_{S}=  0.35$ at $x= 10^{-5}$ which should be contrasted
with the constant GBW value, $\lambda_{GBW}=0.29$. 
An analysis of BFKL evolution near the saturation boundary
found  that the exponent $\lambda_S$ 
is nearly constant as a function of $\log(1/x)$ in the HERA 
kinematic window~\cite{Triant}.

\section{Nuclear Dependence}

\begin{figure}[t]
\includegraphics[height=8cm,width=9cm]{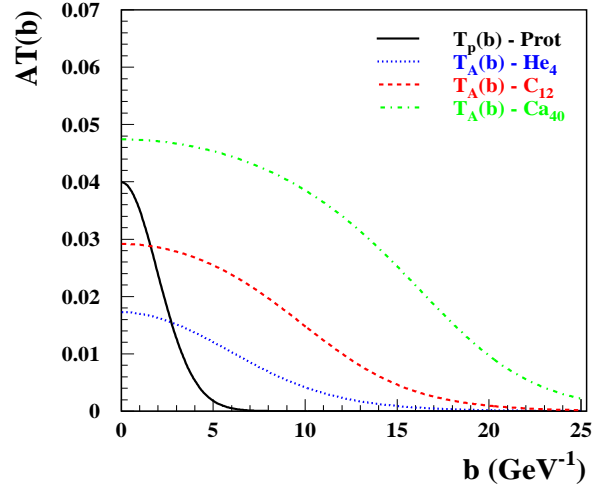}
\caption{\label{fig:tfunc}
  The transverse density $A\,T_A(b)$  for
  several light nuclei compared to the  proton transverse profile, $T_p(b)$. 
}
\end{figure}

Now that impact parameter distribution of the proton is taken into account,
the  dipole picture 
provides a framework for modeling  nuclear phenomena in DIS. 
The distribution of nucleons in the nucleus $\rho_{A}(\vec{r})$  
is parametrized according to the Woods-Saxon distribution \cite{Bohr}
\begin{equation}
   \rho_A(\vec{r}) = \frac{N}{
                           \exp\left( \frac{(r-R_A)
                                           }{ \delta
                                           }
                               \right)
                      + 1 } \;,
\end{equation}
where $A$ denotes the number of nucleons in the nucleus,  $\delta =
0.54 \, \mbox{fm}$, $R_A = (1.12\, \mbox{fm})\, A^{1/3}  - 
(0.86\, \mbox{fm})\, A^{-1/3}$ 
and $N$ is adjusted to normalize the distribution to one
\begin{equation}
   \int d^3\vec{r} \,\rho_A(\vec{r}) = 1 \; .
\end{equation}
The transverse distribution is defined as
\begin{eqnarray}
T_A({\bf b}) = \int_{-\infty}^{+\infty} dz \rho_A (\sqrt{ {\bf b^2} +
z^2} ).
\end{eqnarray}
Figure~\ref{fig:tfunc} compares the proton shape $T_p(b)$ with
the transverse density $T_A(b)$ 
 for several light nuclei.

\begin{figure*}[htp]
\includegraphics[height=8.5cm,width=8.5cm]{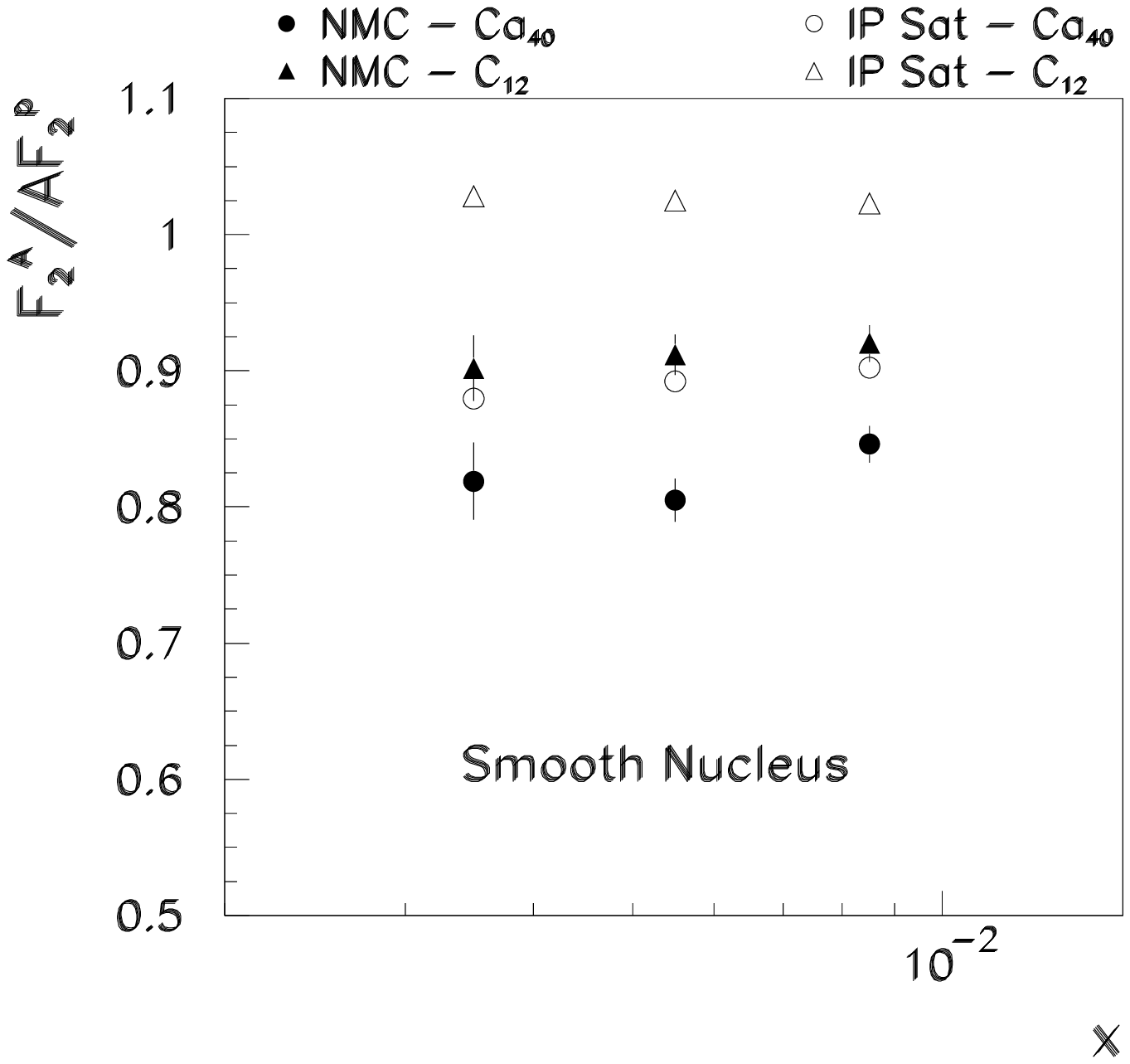}
\includegraphics[height=8.5cm,width=8.5cm]{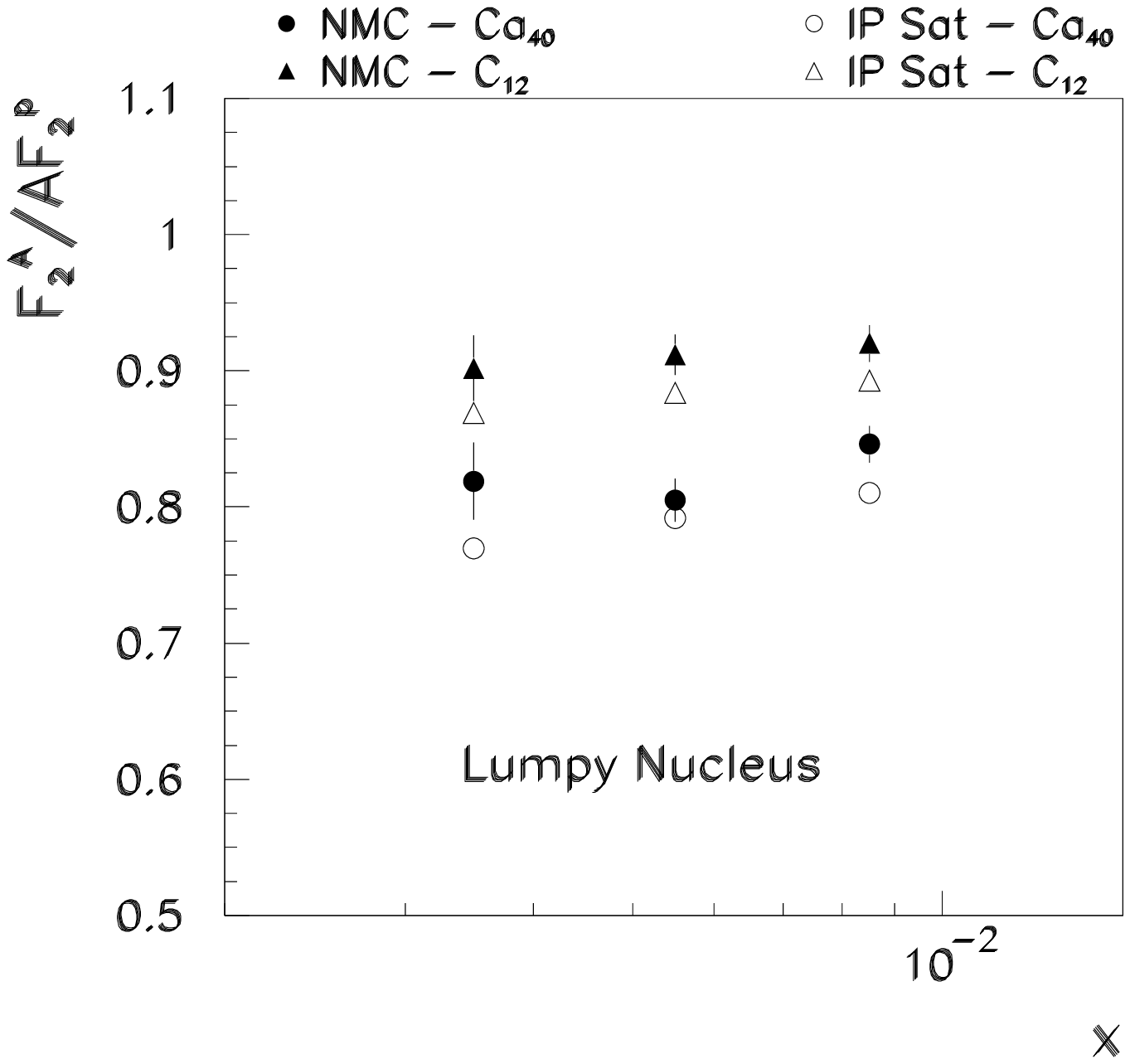}
\caption{\label{fig:nmc}
  Shadowing in DIS: a comparison of model predictions (open points)  
  to data (full points) with the smooth and lumpy nucleus assumptions. The NMC
  measurements\cite{NMC.dat} were performed   at
  $Q^2=0.74,1.20,1.70$ GeV$^2$ for 
  Carbon and at $Q^2=0.60,0.94,1.40$ GeV$^2$ for Calcium
  (in order of increasing $x$ values).
}
\end{figure*}

For a first estimate of  nuclear effects we assume that
the gluons are distributed uniformly throughout the nucleus. This
assumption is realized by replacing the proton shape 
in Eq.~\ref{eq:dipgm} by the integrated Woods-Saxon distribution $T_A(b)$ scaled by the
number of nucleons, 
\begin{equation}
    \frac{
                  d \sigma_{
                               q\bar{q}
                               }
                 } {
                  d^2b
                }  = 2  \left[
                       1 - \exp \left(
                                          - r^2 
\frac{\pi^2}{2N_c} \alpha_s(\mu^2)  x g(x, \mu^2)\, A\, T_A( b)
                                    \right)
                          \right] \; .
\label{eq:dilut}
\end{equation}
 This naive assumption, called the {\it smooth}
 nucleus assumption below, 
cannot be true since Fig.~\ref{fig:tfunc} shows 
that the  center of  the proton is  on average more dense than the
carbon nucleus. 
Nevertheless it is instructive to evaluate $F_2^A/AF_2^p$
for this case  and to  compare it with NMC
measurements of shadowing  at small-$x$~\cite{NMC.dat}, see
Fig.~\ref{fig:nmc}.  
The figure shows that the {\it smooth} nucleus assumption underestimates 
nuclear shadowing considerably as can be anticipated
by comparing the densities in Fig.~\ref{fig:tfunc}.

The proton profile $T_p(b)$  indicates  that the
gluons are distributed within individual nucleons with a radius
of $\approx 0.5\,\mbox{fm}$.
The nucleons are then distributed within the nucleus according to a
Woods-Saxon distribution.
Therefore the gluons form a {\it lumpy} distribution
within the nucleus.  
To account for this correlation among the gluons
we generalize the model in the following way.
 The probability of a given configuration
of nucleons  within the nucleus
$\left\{ 
({\bf b_1}, z_1),  ({\bf b_2}, z_2), \cdots, ({\bf b_A},z_A) \right\} $
is
\begin{equation}
     P_{ \left\{ ({\bf b_1}, z_1),  ({\bf b_2}, z_2), \cdots, ({\bf b_A},z_A)
\right\} } = \Pi_{i=1}^{A} \rho({\bf b_i}, z_i) .
\end{equation}
For a given configuration of nucleons the probability that the dipole
scatters at 
impact parameter $\bf b$ is
\begin{widetext}
\begin{equation}
        \left(  \frac{ d\sigma_{q\bar{q} } } {d^2b} 
        \right)_{  
            \left\{ ({\bf b_1}, z_1),  ({\bf b_2}, z_2), \cdots, 
                   ({\bf b_A},z_A) 
            \right\} 
                } =
  2 \left[ 1 - \exp \left( -\frac{\pi^2}{2N_C}r^2 \alpha_s(\mu^2)x g(x, \mu^2) 
\sum_{i=1}^A  T_p ({\bf b -  b_i})   \right) \right].
\end{equation}
Now we may average over all possible configurations of nucleons to
determine the
cross section
\begin{equation}
       \frac{ d\sigma_{q\bar{q}}} {d^2b}  = \int d^2b_1\,dz_1 \cdots
d^2b_A\,dz_A
             \, P_{ \left\{ ({\bf b_1}, z_1),  ({\bf b_2}, z_2), \cdots, ({\bf
b_A},z_A) \right\} } \,
\left(  \frac{d\sigma_{q\bar{q}} } {d^2b} \right)_{
\left\{ ({\bf b_1}, z_1),  ({\bf b_2}, z_2), \cdots ({\bf b_A},z_A) \right\} }  \; .
\end{equation}
Noticing that the integrals factorize we have
\begin{eqnarray}
   \frac{ d\sigma_{q\bar{q}} }{d^2b}  &=&  2[1  -  (1- I)^A ] \\
                     I  &\equiv& 
\int d^2 {b^{'}}  T_A({\bf b +  b^{'}}) 
\left[
       1- \exp\left( -\frac{\pi^2}{2N_C} r^2 \alpha_s(\mu^2)x g(x, \mu^2) 
T_p({\bf b^{'}})  \right)
\right] \; .
\end{eqnarray}
\end{widetext}

To simplify this further, we notice that the integral  over $\bf b^{'}$ is
 different from zero only over the size of the proton which is small
compared to
the nucleus even for the deuteron.  Thus  $T_A({\bf b +  b^{'}})\approx
T_A({\bf b})$.
Using the definition of  the total cross section of the $q\bar{q}$ pair on
the proton  $\sigma_{q \bar{q}}^p$ (Eq. \ref{eq:diptot} and \ref{eq:dipgm})
  we have 
\begin{eqnarray}
  \frac{ d\sigma_{q\bar{q}} } {d^2b}  & =& 
 2 \left[1  -  (1- \frac{ T_A({\bf b}) }{2}
                    \sigma_{q \bar{q}}^p
                )^{A}
   \right]   \; .
\label{er:glauber}
\end{eqnarray}
Only if A is large and  $r$ is small  do we recover the {\it
  smooth} nucleus  formula of  Eq.~\ref{eq:dilut}.  
  Similar considerations were discussed in the 
  context of setting up initial conditions for 
  classical QCD simulations of nucleus-nucleus collisions \cite{KNV2}.
  Figure~\ref{fig:nmc} 
shows that   the {\it lumpy} nucleus model
slightly 
overestimates the observed shadowing. 
However, good overall agreement with data  clearly
indicates that the IP Saturation model can be used to study
of nuclear effects. 

To estimate the saturation scale for heavy nuclei we follow 
the same procedure as for proton. 
We set the lumpy nucleus S-matrix element squared, $|S|^2 =  (1 - \frac{T_{A}({\bf b}=0)}{2} \sigma_{q\bar{q}}^p)^{2A} $ equal to $e^{-1}$. This  is
an implicit equation for the saturation radius, $r_{S}$. 
Once $r_{S}$ is determined numerically the saturation scale 
is determined from the relation, $Q_{S}^2 = 2/r_{S}^2$.

\begin{figure}[htp]
\includegraphics[height=9cm,width=9cm]{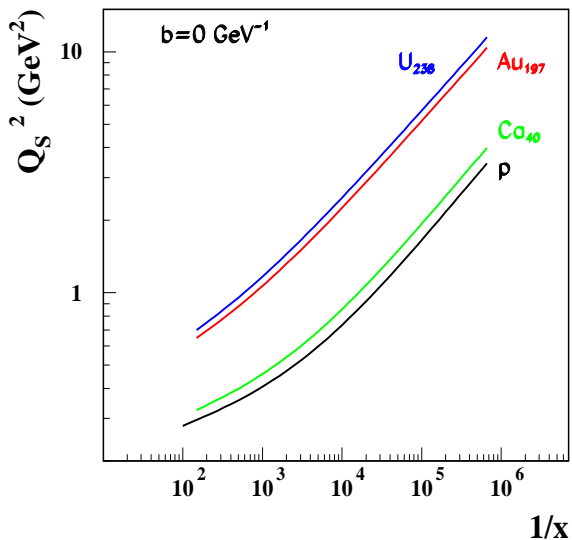}
\caption{\label{fig:q2sxa}
  The quark saturation scale $(Q_{s}^2)_q$ at $b=0$ 
  as a function of $x$ for various nuclei .  The
  gluon saturation scale is 
$(Q_{s}^2)_g = \frac{N_c}{C_F} (Q_{s}^2)_q  \approx 2.25 \,(Q_s^2)_q$.
}
\end{figure}

Figure~\ref{fig:q2sxa} shows the saturation scale  for various
nuclei calculated in this fashion. It is
interesting to observe that the saturation scale is not increasing
like $A^{1/3}$. Rather, for  
Au the nuclear enhancement factor is only a factor of $\approx2$ at
$x=10^{-2}$ and  $\approx3$ at $x=10^{-4}$. 
When comparing the saturation scale to other estimates it is 
important to remember that this is the saturation scale for quarks. 
The saturation scale for gluons is 
$(Q_{s}^2)_g = \frac{N_c}{C_F} (Q_{s}^2)_q  \approx 2.25\,
(Q_s^2)_q$~\cite{am-qm}.  
Thus for RHIC conditions with $x\approx10^{-2}$, $(Q_{s}^2)_g$ is
$\approx 1.3 \, \mbox{GeV}^2$ 
which is slightly smaller than 
was estimated previously by analyzing RHIC multiplicity
data~\cite{am-qm,KZN,KNV,KZL}.  

\section{Summary}

We developed an Impact Parameter dipole saturation model which
describes HERA the total $\gamma^*p$ cross section  and differential 
diffractive $J/\psi$
with high precision.    
The new feature of the model is an impact parameter dependent
dipole cross-section which respects unitarity constraints.
The impact parameter distribution is determined 
by measured $J/\psi$ $t$-distributions to a large extent.
For the functional form of the dipole cross-section we adopt  the
Glauber-Mueller formula~\cite{ref.GlMu} which can also be obtained
within  the
McLerran-Venugopalan model~\cite{ref.McLV}. 
    
The parameters of the dipole cross-section  are determined
from a fit to HERA data.  
The fit depends 
on the specific form of photon and the $J/\psi$ wave function. 
At short distances,
the photon wave-function is well known from QED.
For the $J/\psi$ wave-function we proposed several ansatzes and
verified that  the dipole cross-section was almost
independent of these alternatives. 

QCD evolution effects are taken into account by evolving the gluon
structure function with the lowest order DGLAP equations. DGLAP evolution is
essential  in contrast to the GBW model.

In the dipole  approach only the gluon structure function is modeled.
All sea-quark contributions to $F_2$ are obtained from the
photon wave-function and the subsequent  
dipole-gluon scattering. 
The validity of this approximation 
is corroborated by the excellent agreement between the
predicted and observed open charm contribution to $F_2$. 

The impact parameter dependence of the dipole cross section determines
the properties of the  $J/\psi$ total and diffractive cross-sections.
The saturation ansatz for the dipole cross section
predicts diffractive dips in the $t$ distribution which are
not necessarily predicted by an ansatz without saturation.
For $J/\psi$ production the
diffractive dips 
appear in a region of $t$ which is not covered by data but
which is
accessible to future measurements. In addition, the saturation ansatz predicts 
a slight increase of the diffractive slope $B$,
which reflects a small increase in the proton size at small $x$. The change of
the slope $B$ with energy, called shrinkage, provides 
information about the interplay of saturation and evolution phenomena.
Since  shrinkage was observed for several vector-meson species 
and since considerably more data exists than was used in the present paper, 
further study of diffractive vector meson production 
and their $t$-distributions is warranted.
We also point out  that additional
measurements of vector-meson production in a extended $t$-range
would provide crucial information about the gluon density.

With an accurate determination of the S-matrix element we evaluated
the saturation scale $Q^2_S(x)$ -- the scale at which  multiple
collisions becomes important. The $Q_S^2$ determined in the model is
approximately half of the GBW value and is only weakly
dependent on the specific form of the dipole 
cross-section. 
 
In the low $Q^2$ region up to $\sim 4$ GeV$^2$, saturation dynamics
contributes significantly to the observed cross-sections. The
saturated state is  at least partially perturbative since it is
 frequently initiated by small dipoles.
Above $Q^2\sim 10$ GeV$^2$ saturation plays  a secondary
role. Dipoles are very small and although they are strong gluon
emitters the gluon phase-space density remains small. 

In order to isolate the effects of saturation, we also studied a
dipole model which is not unitarized . This IP
Non-Saturated model also obtained a good description
of the data.  However, even with the non-saturated 
model the fit returned large gluon densities and 
unitarity corrections became important in the
same   region of $x$ and $Q^2$. 
Thus, the saturation ansatz is the more consistent approach.

The extension of the model to nuclei is fairly straightforward.
We evaluated the nuclear cross-section with two
different ansatzes for the distribution of gluons within the nucleus and
compared the results to data on nuclear shadowing at low $x$. We found
that the {\it lumpy} nucleus ansatz, in which gluons are
distributed around the nucleons in a similar manner as around a free
proton, leads to a good 
description of the data. We then used this ansatz to evaluate the
saturation scale in heavy nuclei and found that $Q^2_S$
  increases more slowly then $A^{1/3}$.
Nevertheless, for Gold and Uranium $Q^2_S$ is enhanced by factor of
$\sim 2$  when compared to the  proton. 
 For RHIC
conditions the gluon saturation scale was found to be 
$Q_{s}^2 \approx 1.3 \,\mbox{GeV}^{2}$, which is somewhat smaller
than was estimated previously from RHIC multiplicity data~\cite{am-qm}. 

The success of the model suggests that the saturated state 
known  as the Color Glass Condensate
plays an important role at  HERA.
Saturation is particularly important in the small $x$ and low 
$Q^2$ region where  
the $\gamma^* p$ total inclusive cross-section exhibits the same
universal rate of growth as all known  hadronic cross-sections. 
Revealing the Color Glass Condensate experimentally
would represent a fundamental advance towards an   
understanding of the high energy limit of QCD.

\section{Acknowledgment}

We have profited from helpful discussions with many colleagues:
A. Caldwell, J. Jalilian-Marian, D. Kharzeev,  D. N.
Triantafyllopoulos  and R. Venugopalan. Our 
special thanks go to  A. H. Mueller and L. McLerran. 
One of us (H.K.) thanks the Physics Department of Brookhaven
National Laboratory and the Department of Physics of Columbia
University for hospitality. Special thanks go to the Nevis Laboratories of
Columbia University. 

This work was partially supported by the National Science
Foundation grant NSF PHY 00-98826 and by the 
U.S. Department of Energy grant DE-AC02-98CH10886. 

\appendix
\section{The $J/\psi$ Wave-Function}
\label{wave-append}

The discussion here closely follows \cite{DKP}  and
uses some of the concepts of \cite{FS}.
The light cone wave-function for $J/\psi$ 
can be written as an expansion in the Fock space,
\begin{eqnarray}
   \left|V\right> = \tilde{\psi}_{c\bar{c}} \left|c\bar{c}\right>
        + \tilde{\psi}_{q\bar{q}g} \left|q\bar{q}g\right> + ...  .
\end{eqnarray}
Below we neglect higher Fock components in the 
wave-function and treat the meson as a predominantly
a $c\bar{c}$ pair.  Restoring indices, the $J/\psi$ state
is given by 
\begin{widetext}
\begin{eqnarray}
   \left|V^{\lambda}\right> = 
  \int \frac{d^2k dz}{16 \pi^3}
  (\tilde{\psi}_V)^{h\bar{h}}_{\lambda}({\bf k},z) \frac{\delta_{ab}}{\sqrt{N_{c}}}
  \left| c^{a}({\bf k},z,h) \,\bar{c}^{b}(-{\bf k},1-z,\bar{h}) \right>
\end{eqnarray}
\end{widetext}
Here $\lambda=0,\pm1$ denotes the polarization of the 
vector meson, $h,\bar{h}=\pm$ denote the helicities of
the c-quark and anti-c-quark respectively, and $a,b$ are the color indices of
the c-quark and anti-c-quark.  
Here and below we strictly adhere to the light-cone conventions
of Ref. \cite{ref.BL}. 

The $J/\psi$ wave-function is modeled after the 
photon wave-function. 
The longitudinally polarized vector meson ($\lambda = 0$) is 
\begin{eqnarray}
\label{lc1}
    (\tilde{\psi}_V)^{h\bar{h}}_{0}(k,z) &=&
\frac{ \delta_{h+\bar{h}-} + \delta_{h-\bar{h}+} }
{\sqrt{2}} \tilde{\phi}_L(k,z) \, ,
\end{eqnarray}
while the two transverse polarizations ($\lambda=\pm1$) are given by
\begin{widetext}
\begin{eqnarray}
\label{lc2}
(\tilde{\psi}_V)^{h\bar{h}}_{+1}(k,z) &=& \left[+ \frac{k}{m_c} e^{+i\varphi_k} 
(z \delta_{h+\bar{h}-} - (1-z) \delta_{h-\bar{h}+}) + \delta_{h+\bar{h}+} \right] 
\tilde{\phi}_{T} (k,z) \nonumber \\
(\tilde{\psi}_V)^{h\bar{h}}_{-1}(k,z) &=&\left[ -\frac{k}{m_c} e^{-i\varphi_k}  
( (1-z) \delta_{h+\bar{h}-} - z \delta_{h-\bar{h}+}) + \delta_{h-\bar{h}-} \right] \tilde{\phi}_T (k,z) ,
\end{eqnarray}
\end{widetext}
where $k\,e^{i\varphi_k} = k_x + i k_y$.
Once the wave-function in momentum space is specified then Fourier transforms
return the  wave-function in coordinate  space
\begin{equation}
     (\psi_V)^{h \bar {h}}_\lambda (r,z) = \int \frac{d^2k}{(2\pi)^2} e^{i\bf{k} \cdot \bf{r} }  
     (\tilde{\psi}_V)^{h \bar{h}}_\lambda (k, z) \; .
\end{equation}
Note under Fourier transform $\pm k\, e^{\pm i \varphi_k}\, \tilde{\phi}(k,z)$ 
becomes $\mp i\, e^{\pm i \varphi_r} \,\partial_r\phi(r,z)$.
The overlap with the photon 
wave-function is then
\begin{widetext}
\begin{eqnarray}
\label{olapap}
    (\gamma|V)_{L}\equiv \sum_{h\bar{h}} (\psi_{\gamma}^{*})_{0}^{h\bar{h}} (\psi_{V})^{h\bar{h}}_{0} &=&  e_c \sqrt{2 N_c} \left[ 
    2 z (1-z) Q \,\frac{ K_0(\epsilon r) } {2 \pi}\, \phi_{L} \right] \nonumber \\
   (\gamma|V)_{T}\equiv \frac{1}{2} \sum_{\stackrel{\scriptstyle h\bar{h}}{\lambda=\pm1}} (\psi_{\gamma}^{*})_\lambda^{h\bar{h}} 
      (\psi_V)_\lambda^{h\bar{h}} &=& e_{c} \sqrt{2 N_{c}} \left[ ( z^2 + (1-z)^2 ) 
       (-\frac{1}{m_c} \frac{\;\partial \phi_T}{\partial r}) \frac{\epsilon K_1(\epsilon r) } {2\pi}  + \phi_T \, m_c \frac{K_{0} (\epsilon r) } {2\pi} \right] \; .
\end{eqnarray}
\end{widetext}

Two inputs are used to constrain the wave-function
$\tilde{\phi}_{L,T}$. The first  
input is the normalization of the wave-function, 
$\left<V^{\lambda'}\right.\left|V^{\lambda}\right> = 
(2\pi)^3 2 p^{+}\delta (p'^+ - p^+)\delta^2({\bf p'} - {\bf p})
\delta^{\lambda'\lambda}$. 
This leads to the requirement that
\begin{eqnarray}
\label{norm}
  1 = \int \frac{d^2k dz}{16 \pi^3} \sum_{h\bar{h}} 
  \left|(\tilde{\psi}_V)^{h\bar{h}}_\lambda(k,z) \right|^2  ,
\end{eqnarray}
or 
\begin{equation}
    1 =  \int \frac{d^2k dz}{16 \pi^3} \left| \tilde{\phi}_L(k,z) \right|^2 \\
\end{equation}
for  $\tilde{\phi}_{L}$ and 
\begin{equation}
  1 =  \int \frac{d^2k dz}{16 \pi^3}  \left\{(z^2 + (1-z)^2) \frac{k^2}{m_c^2} + 1\right\}
                \left| \tilde{\phi}_T(k,z) \right|^2
\end{equation}
for $\tilde{\phi}_{T}$ . 
Note that the formula (B19b) in \cite{DKP}  is incorrect
by a factor of two for the transverse wave-function.
The second input is the leptonic decay width of the vector meson. The
decay width of the meson is given by 
\begin{eqnarray}
   \Gamma_{V} &=& \frac{4\pi \alpha_{em}^2}{3} \frac{f_{V}^2}{M_V} \;, 
\end{eqnarray}
where $f_{V}$ is given by
\begin{eqnarray}
 e M_{V} f_{V} \epsilon^{\mu}_\lambda &=& \left< 0\right| J_{em}^{\mu} \left| V^{\lambda}
   \right> \; .
\end{eqnarray}
In this formula $\epsilon^{\mu}_{\lambda}=(\epsilon^{+}, \epsilon^{-}, \epsilon_x, \epsilon_y)$ is the polarization vector 
of the meson and $e$ is the charge of the positron.  
$\epsilon^{\mu}_{\pm 1}= 1/\sqrt{2} (0, 0, 1, \pm i)$  
for transversely polarized mesons and 
$\epsilon^{\mu}_{0}= (\frac{p^{+}}{M_V}, -\frac{M_V}{p^+},0,0)$ 
for the longitudinally polarized mesons. Evaluating this 
matrix element in light cone perturbation with the 
conventions of \cite{ref.BL} gives
\begin{widetext}
\begin{eqnarray}
\label{decay}
     f_{V} &=& \hat{e_c} \sqrt{2 N_c}\, 2 
     \int \frac{d^2k dz}{16 \pi^3} \tilde{\phi}_{L} (k,z) \nonumber \\
     f_{V} &=& \hat{e_c} \sqrt{2 N_c}\,  \frac{m_{c}}{M_V} 
     \int \frac{d^2k dz}{16 \pi^3}  
     \left\{(z^2 + (1-z)^2) \frac{k^2}{m_c^2} + 1\right\}
     \frac{ \tilde{\phi}_{T} (k,z) }{z(1-z)} \; , 
\end{eqnarray}   
\end{widetext}
where $\hat{e_c}=+\frac{2}{3}$  (in contrast to 
$e_c=\sqrt{4 \pi \alpha_{em}}\hat{e_c}$\,)
is the charge in units of the positron charge.  

To go further, we must specify the form of the wave-function.  
We have studied three model wave-functions, GAUS-RF, CORNELL, and
GAUS-LC. Two of the models (GAUS-RF and CORNELL) are 
motivated by the expected long distance behavior of the 
wave-function.  One of models (GAUS-LC) is motivated by the expected 
short distance behavior of the wave-function.

For large distances, and for truly heavy vector mesons, the 
non-relativistic wave-function  should give a good description
of the light cone wave-function. 
To make the correspondence 
between the light cone and non-relativistic wave-functions 
functions we repeat the discussion in Ref. \cite{FS}.
In the non relativistic limit, terms proportional to
$\left(\frac{k_t}{m_c}\right)^2$ can be dropped and the  
the normalization condition Eq.\,\ref{norm} is identical for the longitudinal
and transverse wave-functions. 
Simple kinematic relations give the invariant mass 
of the the quark antiquark pair  
$M_{V}^2 = (p_{1} + p_{2})^2 = 
\frac{ m_{c}^2 + k_{t}^2 }{  z (1 -z) }$. This
relation assumes that the quark and antiquark are on shell, which
is true for non-relativistic quarks to order $\sim \left(\frac{k_t}{m_c}\right)^2$. 
In the rest frame of the meson (where the 
quarks have momentum $\vec{p}$ and $-\vec{p}\,$)\, the invariant
mass is $(p_{1} + p_{2})^2 = 4 (p^2 + m_{c}^2)$.
Comparing these terms gives the identification
\begin{eqnarray}
    p^2 &\rightarrow&  \frac{k_{t}^2 + (2z-1)^2 m_c^2 }{4 z(1-z)}  \\
    d^3p &\rightarrow&  \frac{\sqrt{k_{t}^2 + m_c^2}}{4 \left[z(1-z)\right]^{3/2} } d^2k_{t} dz  \; .
\end{eqnarray}
Requiring probability conservation per unit phase space
\begin{equation}
     \frac{d^3p}{(2\pi)^3} \left|\tilde{\psi}_{NR}(p)\right|^2 = 
     \frac{d^2k_t dz}{16 \pi^3} \left|\tilde{\phi}_{L,T} (k_t,z)\right|^2 \; ,
\end{equation}
leads to the identification
\begin{widetext}
\begin{equation}
\label{lc3}
   \tilde{\phi}_{L,T}(k_t,z) = 
   N_{L,T} \left( \frac{ k_t^2 + m^2_c  } { 4 z^3 (1-z)^3}  \right)^{\frac{1}{4}} 
   \tilde{\psi}_{NR}\left(
   p= \sqrt{
            \frac{ k_t^2 + (2z -1)^2 m_c^2}
                 {  4 z (1 -z) }
           }
           \right) \; .
\end{equation}
\end{widetext}
Here $\tilde{\psi}_{NR}(p)$ is the non-relativistic wave-function 
in momentum space
\begin{equation}
   \psi_{NR}(\vec{x}) = \int \frac{d^3p}{(2\pi)^3} e^{i\vec{p}\cdot\vec{x} }
                          \tilde{\psi}_{NR}(\vec{p}) \; .
\end{equation}

For the two long distance models (GAUS-RF and CORNELL) 
$\tilde{\psi}_{NR}(p)$ is specified.
Then with the form of $\tilde{\phi}_{L,T}$ specified, the
non-relativistic motivation for the light cone 
wave-function is abandoned and
$N_{L,T}$ are constants adjusted to satisfy the
normalization and decay conditions (Eqs. \ref{norm} and \ref{decay}) 
including the
terms proportional to $\left(k_t/m_c\right)^2$. 
Terms proportional to $\left(k_t/m_c\right)^2$ mock up 
relativistic effects and constitute $\approx 25\%$ of the 
total elastic diffractive cross section.
For the short distance model (GAUS-LC) $\tilde{\phi}_{L,T}$ is
specified directly.
For the numerical results quoted below we 
take $M_V = 3.096\;$GeV and $m_c = 1.3\;$GeV.

\noindent \underline{GAUS-RF}. 
In this model we take 
\begin{equation}
\tilde{\psi}_{NR}(p) = N_{L,T}
\exp\left(-\frac{p^{2}}{\omega_{L,T}^2}\right),
\end{equation} 
where $\omega_{L,T}$ and $N_{L,T}$ are adjusted to satisfy 
Eq. \ref{decay} for the decay width and  Eq. \ref{norm} for the
norm. These parameters are $N_{L} =1.74 $ GeV$^{-3/2}$, $\omega_{L}=0.66$
GeV  and  
$N_{T}=1.86$ GeV$^{-3/2}$ and $\omega_{T}=0.61$ GeV. 

\noindent \underline{CORNELL}. 
In this model we take the form of $\tilde{\psi}_{NR}(p)$ from the
wave-function  
of a non-relativistic potential model of $J/\psi$. Using a 
shooting method \cite{NR} 
we solve the radial Schr\"odinger equation for the Cornell potential to
find $\psi_{NR}(\vec{x})$.
The parameters of the  
potential are taken from Ref. \cite{Eichten}.  Note, the constituent quark
mass of the potential model ($m_c=1.8\;\mbox{GeV}$) 
differs from the current quark mass 
($m_c=1.3\;\mbox{GeV}$) in Eqs.
\ref{lc1}, \ref{lc2}, \ref{lc3} for the the light cone wave-function. 
In this model the decay 
condition Eq. \ref{decay} is not satisfied. 

%

\noindent \underline{GAUS-LC}. 
We now turn to the short distance model GAUS-LC. 
At short distances the light cone vector meson wave-function is
proportional to  
$\propto z(1-z)$ \cite{ref.BL,FS}. 
This motivates the following ansatz for the wave-function 
\begin{equation}
\tilde{\phi}_{L,T}(r,z) = N_{L,T} \,z (1 -z)\, \exp(- \frac{k^2}{\omega^2_{L,T}})
\end{equation}
$N_{L,T}$ and $\omega_{L,T}$ are adjusted to reproduce the
normalization condition and the decay width. For $J/\psi$
we find: $N_{L} = 1.21$ GeV$^{-1}$, $\omega_{L}=0.80$ GeV
and $N_{T}=1.63$ GeV$^{-1}$ and
$\omega_{T} =0.58$ GeV.

With the wave-function $\tilde{\phi}_{L,T}(k,z)$ specified the 
wave-function $\phi_{L,T}(r,z)$ and its derivative $\partial_{r} \phi(r,z)$.
are found taking  Fourier transforms.
The final overlap between the photon and vector meson wave-function is given
by Eq. \ref{olapap}.

\end{document}